\documentclass{aastex631}
\usepackage{amsmath}
\graphicspath{{figures/}}

\begin{document}

\title{Neutral Atmospheric Density Measurement Using \textit{Insight}-HXMT Data by Earth Occultation Technique}

\shorttitle{Neutral Atmospheric Density Measurement}
\shortauthors{Xue et al.}

\correspondingauthor{Li Xiao-Bo, Xiong Shao-Lin}
\email{lixb@ihep.ac.cn, xiongsl@ihep.ac.cn}

\author[0000-0001-8664-5085]{Wang-Chen Xue}
\affiliation{Key Laboratory of Particle Astrophysics, Institute of High Energy Physics, Chinese Academy of Sciences, 19B Yuquan Road, Beijing 100049, China}
\affiliation{University of Chinese Academy of Sciences, Chinese Academy of Sciences, Beijing 100049, China}

\author[0000-0003-4585-589X]{Xiao-Bo Li\textsuperscript{*}}
\affiliation{Key Laboratory of Particle Astrophysics, Institute of High Energy Physics, Chinese Academy of Sciences, 19B Yuquan Road, Beijing 100049, China}

\author[0000-0002-4771-7653]{Shao-Lin Xiong\textsuperscript{*}}
\affiliation{Key Laboratory of Particle Astrophysics, Institute of High Energy Physics, Chinese Academy of Sciences, 19B Yuquan Road, Beijing 100049, China}

\author{Yong Chen}
\affiliation{Key Laboratory of Particle Astrophysics, Institute of High Energy Physics, Chinese Academy of Sciences, 19B Yuquan Road, Beijing 100049, China}

\author[0000-0001-5586-1017]{Shuang-Nan Zhang}
\affiliation{Key Laboratory of Particle Astrophysics, Institute of High Energy Physics, Chinese Academy of Sciences, 19B Yuquan Road, Beijing 100049, China}

\author[0000-0003-0274-3396]{Li-Ming Song}
\affiliation{Key Laboratory of Particle Astrophysics, Institute of High Energy Physics, Chinese Academy of Sciences, 19B Yuquan Road, Beijing 100049, China}

\author{Shu Zhang}
\affiliation{Key Laboratory of Particle Astrophysics, Institute of High Energy Physics, Chinese Academy of Sciences, 19B Yuquan Road, Beijing 100049, China}

\author{Ming-Yu Ge}
\affiliation{Key Laboratory of Particle Astrophysics, Institute of High Energy Physics, Chinese Academy of Sciences, 19B Yuquan Road, Beijing 100049, China}

\author[0000-0003-3127-0110]{You-Li Tuo}
\affiliation{Key Laboratory of Particle Astrophysics, Institute of High Energy Physics, Chinese Academy of Sciences, 19B Yuquan Road, Beijing 100049, China}

\author{Hai-Tao Li}
\affiliation{University of Chinese Academy of Sciences, Chinese Academy of Sciences, Beijing 100049, China}
\affiliation{Key Laboratory of Electronics and Information Technology for Space Systems, National Space Science Center, Chinese Academy of Sciences, Beijing, 100190, China}

\author{Dao-Chun Yu}
\affiliation{University of Chinese Academy of Sciences, Chinese Academy of Sciences, Beijing 100049, China}
\affiliation{Key Laboratory of Electronics and Information Technology for Space Systems, National Space Science Center, Chinese Academy of Sciences, Beijing, 100190, China}

\author{Dong-Ya Guo}
\affiliation{Key Laboratory of Particle Astrophysics, Institute of High Energy Physics, Chinese Academy of Sciences, 19B Yuquan Road, Beijing 100049, China}

\author{Jia-Cong Liu}
\affiliation{Key Laboratory of Particle Astrophysics, Institute of High Energy Physics, Chinese Academy of Sciences, 19B Yuquan Road, Beijing 100049, China}

\author{Yan-Qiu Zhang}
\affiliation{Key Laboratory of Particle Astrophysics, Institute of High Energy Physics, Chinese Academy of Sciences, 19B Yuquan Road, Beijing 100049, China}
\affiliation{University of Chinese Academy of Sciences, Chinese Academy of Sciences, Beijing 100049, China}

\author{Chao Zheng}
\affiliation{Key Laboratory of Particle Astrophysics, Institute of High Energy Physics, Chinese Academy of Sciences, 19B Yuquan Road, Beijing 100049, China}
\affiliation{University of Chinese Academy of Sciences, Chinese Academy of Sciences, Beijing 100049, China}

\begin{abstract}

The Earth occultation technique has broad applications in both astronomy and atmospheric density measurements. 
We construct the background model during the occultation of the Crab Nebula observed by the \textit{Insight}-Hard X-ray Modulation Telescope (\textit{Insight}-HXMT) at energies between 6\,keV and 100\,keV. 
We propose a Bayesian atmospheric density retrieval method based on the Earth occultation technique, combining Poisson and Gaussian statistics. 
By modeling the atmospheric attenuation of X-ray photons during the occultation, we simultaneously retrieved the neutral densities of the atmosphere at different altitude ranges.
Our method considers the correlation of densities between neighboring atmospheric layers and reduces the potential systematic bias to which previous work may be subject. 
Previous analyses based on light curve fitting or spectral fitting also lost some spectral or temporal information of the data. 
In contrast to previous work, the occultation data observed by the three telescopes onboard \textit{Insight}-HXMT is fully used in our analysis, further reducing the statistical error in density retrieval.
We apply our method to cross-check the (semi-)empirical atmospheric models, using 115 sets of occultation data of the Crab Nebula observed by \textit{Insight}-HXMT. We find that the retrieved neutral density is $\sim$10\%, $\sim$20\%, and $\sim$25\% less than the values of the widely used atmospheric model NRLMSISE-00, in the altitude range of 55\textendash80\,km, 80\textendash90\,km, and 90\textendash100\,km, respectively. We also show that the newly released atmospheric model NRLMSIS 2.0 is generally consistent with our density measurements.
\end{abstract}


\section{Introduction} \label{Sect1}

For astronomical satellites working in the low Earth orbit with certain inclinations and precessions, the emission from particular sources can sometimes be attenuated by the atmosphere or occulted by Earth. For all-sky Field of View (FoV) X/$\gamma$-ray monitors, the Earth Occultation Technique (EOT) is first proposed to image or measure the flux and energy spectrum of high-energy astrophysical sources \citep{Zhang1993OccultationImaging, Harmon2002OccultationTechnique, Harmon2004OccultationCatalog, Wilson-Hodge2012OccultationCatalog, Rodi2014OccultationImaging, Singhal2021OccultationCZTI}. For X-ray satellites with a small FoV, the EOT usually serves as a remote sensing measurement of atmospheric density \citep{Determan2007AtmosphericDensity, Katsuda2021AtmosphericDensity, Yu2022AtmosphericDensity1, Yu2022AtmosphericDensity2, Katsuda2022AtmosphericDensity}.

The EOT has long been used in atmospheric measurements at different wavelengths \citep{Smith1990Review}. Using occultations of the Sun, the Stratospheric Aerosol and Gas Experiment II and III (SAGE II and III) measured aerosols in the stratosphere in the infrared band \citep{Mauldin1985SAGEII, Mauldin1993SAGEIII}. Solar occultation can also be used to make density measurements of atmospheric components at ultraviolet wavelengths \citep{Bauer2012NO2}. Based on the occultations of stars, the atmospheric profile can also be retrieved within the ultraviolet-visible and infrared bands \citep{Hays1973StellarOccultation, Kyrola2010GOMOS}. Compared to other wavelengths, the EOT in the X-ray band has the advantage of less mathematical complexity when modeling photon absorption or scattering, which simply involves the interaction between electrons of the atoms and the X-ray photons, without considering the thermodynamics, chemistry, or ionization of atmospheric components \citep{Determan2007AtmosphericDensity}. However, there is a trade-off between the simplicity of absorption or scattering modeling and the capability to distinguish atmospheric constituents. In other words, the molecular components (e.g., N, N$_2$, O, O$_2$) cannot be identified from the X-ray EOT, to which the overall neutral density is sensitive instead. Nonetheless, component decomposition in X-ray EOT is still possible at the atomic level if absorption edges of the molecular components are present in the measured energy band. \citep{Determan2007AtmosphericDensity, Katsuda2021AtmosphericDensity}.

As an environmental parameter of the middle and upper atmosphere, the neutral density plays a critical role in space engineering, for example, the determination of the orbit of the satellite \citep{Storz2005HASDM}, the prediction of the landing point of reentry \citep{Fedele2021PreciseReentry}. Empirical atmospheric models are important approaches to obtaining the neutral density profile. The NRLMSIS series \citep{Hedin1987NRLMSIS, Picone2002NRLMSIS, Emmert2021NRLMSIS} has been validated, revised, and improved many times since its introduction, and has become one of the standard atmospheric models for space research. However, density discrepancies between the NRLMSIS model and the measurements have also been claimed. Previous work has shown that the neutral density measured with X-ray EOT is systematically lower than that of the NRLMSIS model. \cite{Determan2007AtmosphericDensity} first proposed an occultation light curve fitting approach to retrieving neutral densities within the altitude ranges of 70\textendash90\,km and 100\textendash120\,km, which were both lower than the predictions of the NRLMSISE-00 model. \cite{Katsuda2021AtmosphericDensity} introduced an occultation spectrum fitting method to measure the column density along the Line of Sight (LoS), from which the neutral density can be inverted, and found a $\sim$50\% density deficit within the altitude range of 70\textendash110\,km. \cite{Yu2022AtmosphericDensity1, Yu2022AtmosphericDensity2} developed Bayesian methods to model the occultation light curve and spectrum, and the neutral density retrieved within the altitude range of 85\textendash200\,km is also found to be lower than that of the NRLMSISE-00 model.

It is worth using more observations to cross-check whether the neutral density of the NRLMSIS model is systematically greater than that measured with X-ray EOT. Here, we present a method that simultaneously retrieves the neutral densities in different altitude layers, unlike the previous work where the densities of different altitude layers were measured independently, which may introduce some systematic errors, since the LoS actually passes through different altitude layers and the atmospheric attenuation is cumulative. Our method also has the advantage of making nearly full use of the energy information of the data, which could reduce the statistical uncertainty of the retrieved densities. In our analysis, the occultation data of the Crab Nebula acquired with  \textit{Insight}-Hard X-ray Modulation Telescope (\textit{Insight}-HXMT) in 6-100\,keV are used to perform the density retrieval within the altitude ranges of 55\textendash100\,km.

This paper is organized as follows. Section \ref{Sect2} describes the occultation data of the Crab Nebula observed by \textit{Insight}-HXMT and the data reduction, including the background estimation during the occultation. Section \ref{Sect3} presents the details of the Bayesian atmospheric density retrieval method. The analysis results and discussions are given in Section \ref{Sect4}. Section \ref{Sect5} summarizes the paper and provides perspectives.

\section{Observations and Data Reduction} \label{Sect2}

\begin{figure}
    \centering
    \includegraphics[width=0.45\textwidth]{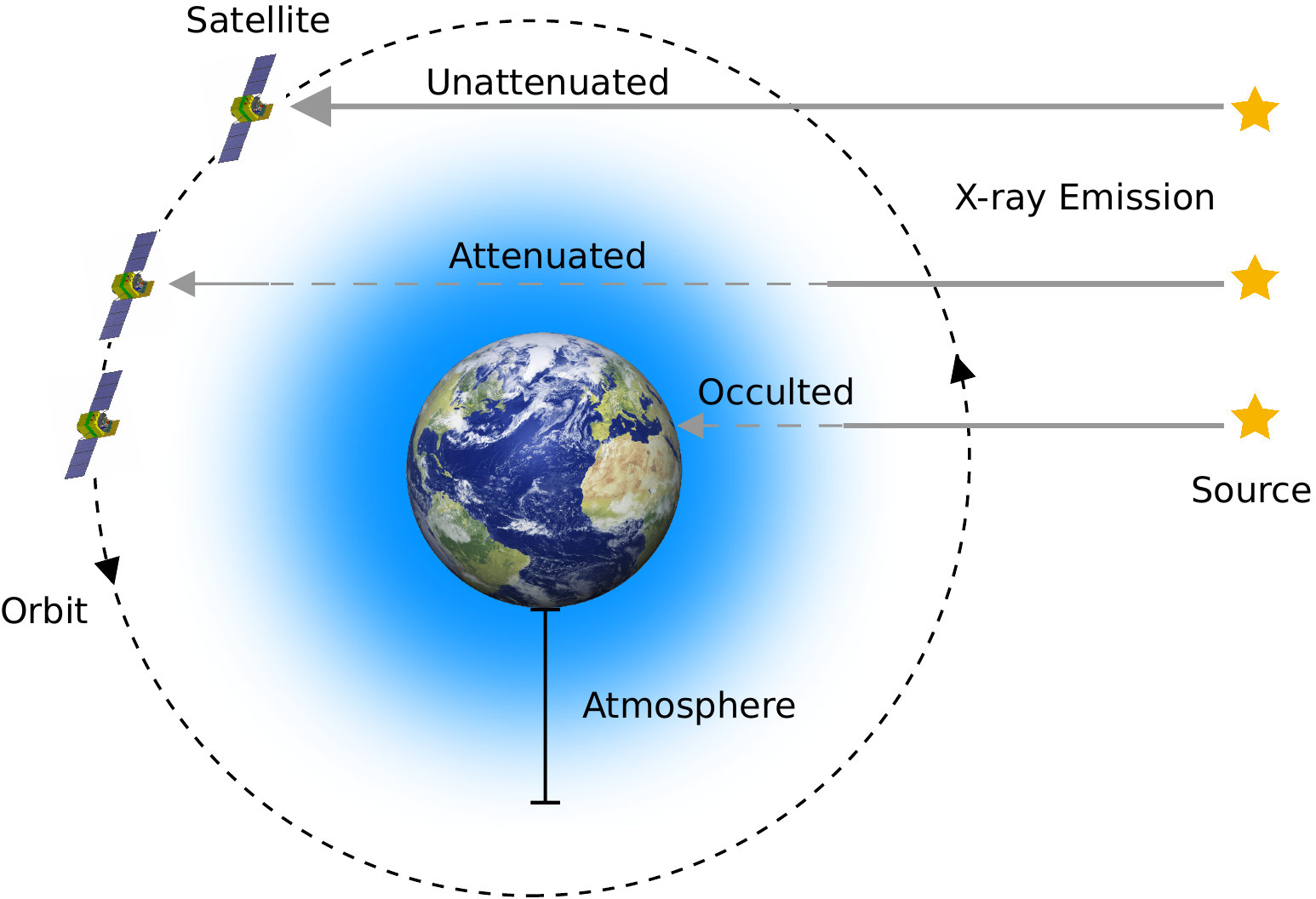}\\
    \caption{Earth occultation of celestial source. The source appear to set (or rise) behind the limb of Earth, due to the satellite motion. The X-ray emission from source will be occulted when the LoS pass through the atmosphere or Earth.}
    \label{Fig1}
\end{figure}

The Earth limb has an aperture of $\sim$134$^{\circ}$ on the celestial sphere for a 550\,km low Earth orbit of an X-ray astronomy satellite such as \textit{Insight}-HXMT. The large Earth-covered areas make it easy for occultation to occur, which typically results in a step-like rising or setting feature in the observed light curves. Figure \ref{Fig1} illustrates the viewing geometry of Earth occultation.

To perform X-ray EOT, there are some restrictions on source selection. The source must be bright enough to obtain a better photon statistic, since a typical time scale of an occultation is only about 30\,s, which needs to be divided into a smaller time bin to be analyzed. Then, the source emission intensity and spectral shape should remain the same during the occultation to avoid confusion between the source variation and photon attenuation caused by the atmosphere. Finally, a point source is preferred for the simplicity of data analysis. Born in a supernova in 1054 AD, the Crab Nebula (along with its pulsar) is a well-known standard candle in the X-ray sky, owing to its brightness, almost constant intensity, and power-law distributed emission spectrum. Numerous X-ray astronomy satellites perform their calibration with the Crab Nebula, including \textit{Insight}-HXMT \citep{Li2020InflightCalibration,Tuo2022ApJS}. The emission area of the Crab Nebula has an angular extent of approximately 1$^\prime$ on the celestial sphere \citep{Madsen2015Crab}. From the satellite's viewpoint at altitude of $\sim$550\,km, this corresponds to an extent of about 0.5\,km near the Earth limb. Although the Crab Nebula is not an ideal point source, this could have little effect on our analysis if a large time bin is chosen to divide the occultation. For example, a time bin of 0.5\,s will produce an altitude difference of $\sim$1.4\,km between two adjacent LoS, which is greater than the extent of the Crab Nebula. Therefore, a time bin of 0.5\,s is adopted in our analysis. 

\subsection{The data reduction of \textit{Insight}-HXMT} \label{Sect2.1}
\textit{Insight}-HXMT, as the first X-ray astronomy satellite in China \citep{Zhang2020Overview}, consists of three scientific payloads: the Low Energy X-ray telescope \citep[LE;][]{Chen2020LE}, the Medium Energy X-ray telescope \citep[ME;][]{Cao2020ME}, and the High Energy X-ray telescope \citep[HE;][]{Liu2020HE}. The observed energy band of LE, ME and HE/NaI covers 1\textendash10\,keV, 10\textendash35\,keV, and 28\textendash250\,keV, respectively. Here, \textit{Insight}-HXMT's pointed observations of the Crab Nebula (ObsID P0111605), which lasted from 2017-11-09T04:03:37 to 2018-08-31T08:50:56 UTC, are reduced to perform the X-ray EOT. The \textit{Insight}-HXMT Data Analysis Software package\footnote{http://hxmten.ihep.ac.cn/} (\texttt{HXMTDAS}) of version 2.04 and the CALibration DataBase (\texttt{CALDB}) of version 2.05 are used to extract high-level scientific products. 

\begin{figure}[htb]
    \centering
    \includegraphics[width=0.5\textwidth]{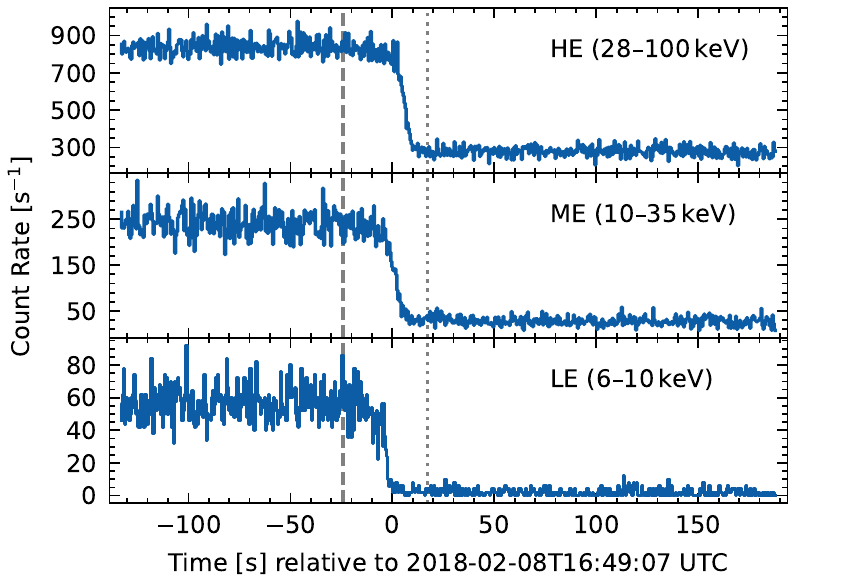}
    \caption{The light curves observed by three X-ray telescopes of \textit{Insight}-HXMT during a setting-type occultation of the Crab Nebula with time bin of 0.5\,s. The emission from the Crab Nebula is first occulted by the atmosphere and then by Earth. The dashed and dotted line indicates the time when the LoS altitude is 150\,km and 40\,km, respectively.}
    \label{Fig2} 
\end{figure}

Here, step-by-step commands to extract occultation data are presented as follows. First, \texttt{lepical}, \texttt{mepical}, and \texttt{hepical} are used to calibrate the photon events of LE, ME, and HE telescopes, respectively. Then, \texttt{lerecon} is applied to the calibrated events of LE telescope to reconstruct two split events and assign the grade, while \texttt{megrade} is used to calculate the grade and the dead time of the ME's calibrated events. With a user-defined criterion, the Good Time Interval (GTI) can be calculated using \texttt{legtigen}, \texttt{megtigen}, and \texttt{hegtigen}. To select the Occultation Time Interval (OTI), the criterion we adopt is \texttt{-10<ELV \&\& ELV<10 \&\& ANG\_DIST<=0.05 \&\& COR>8 \&\& SAA\_FLAG==0 \&\& T\_SAA>=300 \&\& TN\_SAA>=300}. This criterion consists of two parts; the first part \texttt{-10<ELV \&\& ELV<10 \&\& ANG\_DIST<=0.05} is used to select the OTI, while the second part \texttt{COR>8 \&\& SAA\_FLAG==0 \&\& T\_SAA>=300 \&\& TN\_SAA>=300} should ensure that the OTI with high background is excluded. Furthermore, we restrict the OTI to the time interval when the LoS altitude (i.e. the minimum height between the LoS and the ground) is 40--150\,km. For a given OTI, the calibrated event data can be filtered to obtain the occultation data through \texttt{lescreen}, \texttt{mescreen}, and \texttt{hescreen}. The screened event data contain the time (MET) and energy channel (PI) information for each photon event. Finally, the instrumental response file can be generated from \texttt{lerspgen}, \texttt{merspgen}, and \texttt{herspgen}.
The light curves can be extracted through the screened event data with time bin of 0.5\,s.
Figure \ref{Fig2} presents occultation light curves of Crab Nebula acquired with three X-ray telescopes of \textit{Insight}-HXMT.
%In short, the Crab Nebula is a source naturally suited for X-ray EOT.

\subsection{The estimation of background} \label{Sect2.2}
\begin{figure*}[htb]
    \centering
    \includegraphics[width=0.5\textwidth]{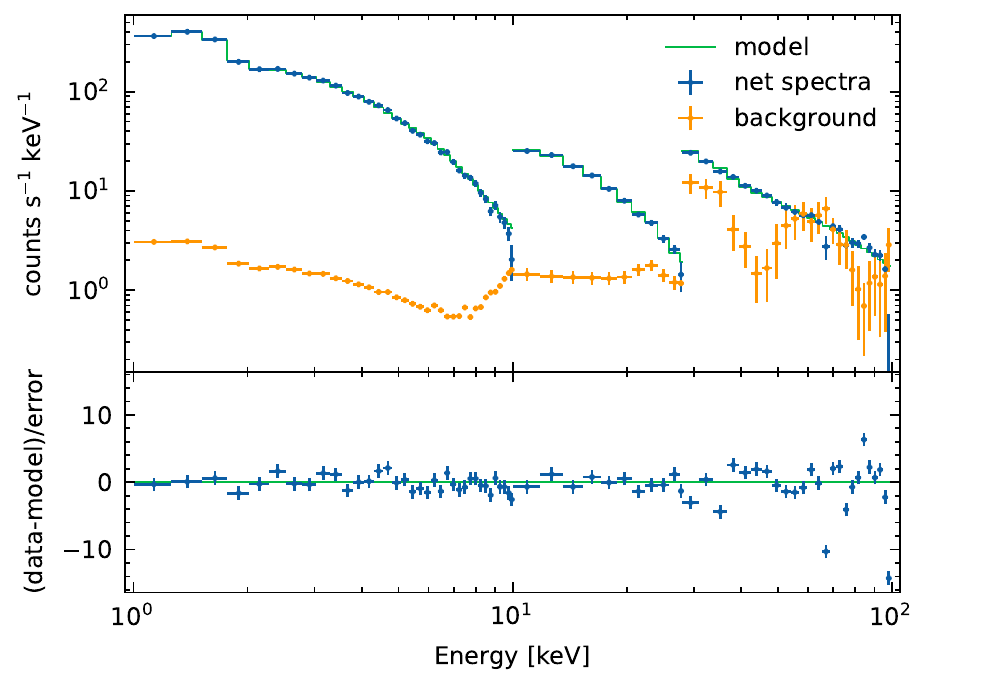}
    \caption{The photon flux of the Crab Nebula observed during a standard GTI (from 2018-02-08T16:28:45 UTC to 2018-02-08T16:29:42 UTC), which is the most adjacent GTI to the occultation data shown in Figure \ref{Fig2}. The green line is the best-fit spectral model, i.e. \texttt{WABS*POWERLAW} in \texttt{XSPEC}, with the parameters of $\mathrm{nH}=0.36$ and $\Gamma=2.10$. The data points in blue are background-subtracted, and the data points in orange are the corresponding background estimated from \texttt{lebkgmap}, \texttt{mebkgmap}, or \texttt{hebkgmap}. For clarity, these data points are binned every 30 channels for LE and ME telescopes, while every 2 channels for HE telescope.}
    \label{Fig:spec} 
\end{figure*}

The background of LE, ME, and HE telescopes has been studied and modeled. 1\textendash6\,keV background of LE is dominated by cosmic X-ray background, and 6\textendash10\,keV background is dominated by charged particles \citep{Liao2020LEBackground}. The main source of the background of ME is charged particles \citep{Guo2020MEBackground}, while the background of HE is mainly due to the interaction of the satellite with various particles \citep{Liao2020HEBackground}. Based on these findings, a standard background estimation procedure has been implemented in \texttt{HXMTDAS}, that is, \texttt{lebkgmap}, \texttt{mebkgmap}, and \texttt{hebkgmap}. Since OTI is usually excluded from standard GTI, the standard background estimation procedure would lose some accuracy when applied to OTI. To obtain a more accurate background estimate, we instead use a time-dependent model to describe the background of OTI. Occultation data are divided into 0.5-s time bins, and this will generate $N$ counting data $\{D_{c,i}\}$ in each energy channel $c$, where $i = 1, \ldots, N$. For each energy channel $c$, the background count rate model is
\begin{equation}
    b_c(t) = \exp{\left(\alpha_c + \beta_c t\right)},
    \label{Eq1}
\end{equation}
where $t$ is the time outside the OTI, then the corresponding light curve model is
\begin{equation}
    m_c(t) = b_c(t) + I_{\mathrm{unocc}}(t) \int_E R(E,c) \, I_0(E) \, \mathrm{d}E,
    \label{Eq2}
\end{equation}
where
\begin{equation}
    I_{\mathrm{unocc}}(t) =
    \begin{cases}
		0, & \text{if source is occulted at time $t$}, \\
        1, & \text{if source is unocculted at time $t$},
    \end{cases}
\end{equation}
$I_0(E)$ is the source flux at energy $E$, which can be obtained from spectral analysis of the most adjacent standard GTI, and $R(E,c)$ is the response matrix of the instrument. Figure \ref{Fig:spec} shows the photon flux of the Crab Nebula observed by the three X-ray telescopes of Insight-HXMT during a standard GTI. Equation (\ref{Eq2}) is used to fit the light curve outside the OTI to obtain the best-fit parameters of Equation (\ref{Eq1}), from which the background estimate of the OTI can be obtained. However, the cosmic X-ray background dominates the 1\textendash6\,keV background of LE telescope, and its variation during the OTI cannot be properly described using Equation (\ref{Eq1}). For simplicity of analysis, we only analyze occultation data of LE telescope within 6\textendash10\,keV (697\textendash1169 channel). For ME and HE telescopes, source attenuation is significant in the energy band of 10\textendash35\,keV (119\textendash546 channel) and 28\textendash100\,keV (8\textendash58 channel), respectively, within which the data will be analyzed.

\begin{figure*}[htb]
    \centering
    \setlength\tabcolsep{0pt}
    \begin{tabular}{cc}
        \includegraphics[width=0.5\textwidth]{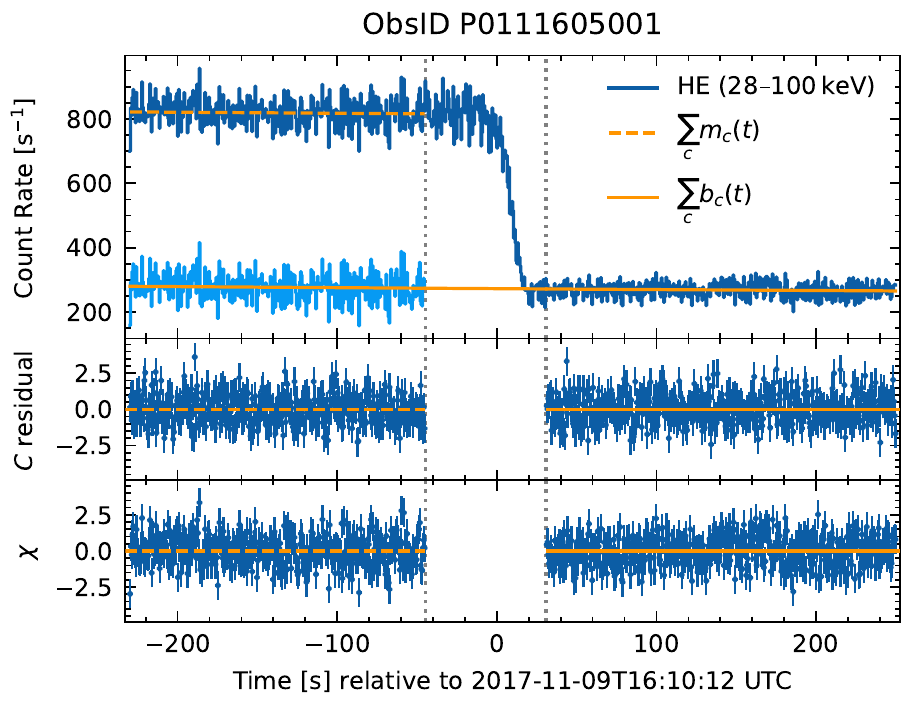}
         & 
        \includegraphics[width=0.5\textwidth]{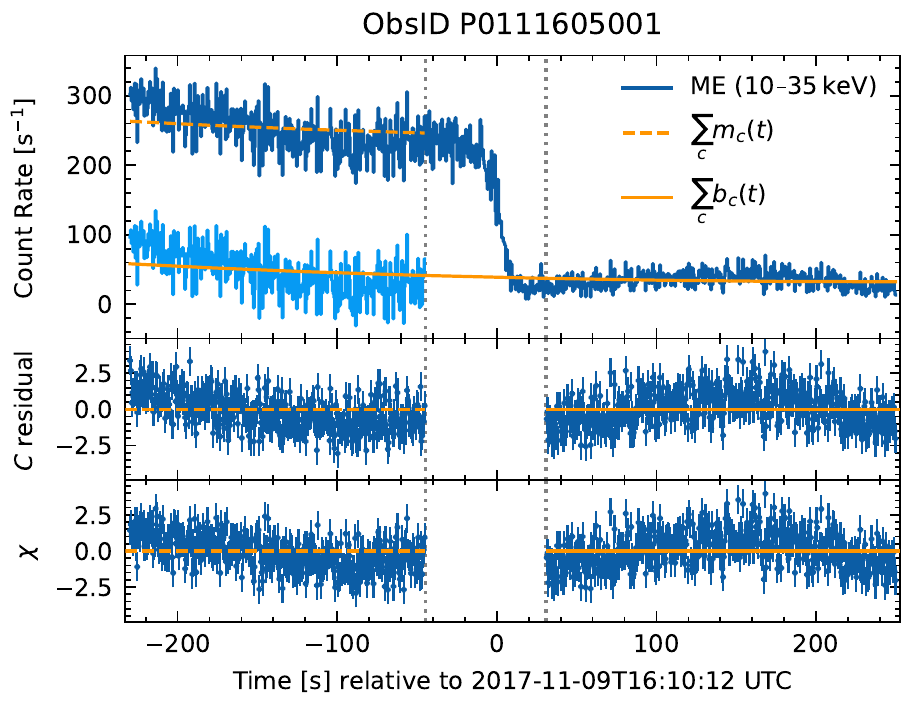}
    \end{tabular}
    \caption{The light curves of one occultation and the corresponding best-fit models for HE and ME telescopes. The reference time is the time when the LoS altitude becomes 90\,km. LE telescope is saturated in this occultation and its data is therefore unavailable. The light blue data points are source flux subtracted, so that the variation of background can be seen more clearly. Two vertical dotted lines indicate the OTI. Two types of residual, $\sum_c \left( C_{c,i} - C_{c,i,\textrm{e}} \right) / \sqrt{\sum_c C_{c,i,\textrm{v}}}$ and $\sum_c \left(D_{c,i} - M_{c,i}\right) / \sqrt{\sum_c M_{c,i}}$ are also plotted to show the fitness of the model. \textit{Left}: The best-fit model of HE light curve passes the goodness-of-fit test, and the corresponding $p$-value is 0.19. \textit{Right}: The best-fit model of ME light curve fails to pass the goodness-of-fit test, due to the fact that the prominent variation of background cannot be well fit by a background model with only two free parameters for each channel.}
    \label{Fig3} 
\end{figure*}

Using the maximum likelihood method, $\{D_{c,i}\}$ outside the OTI is combined to fit Equation (\ref{Eq2}). For counting data, maximizing Poisson likelihood is equivalent to minimizing $C$ statistic \citep{Cash1979Cstat},
\begin{equation}
        C_c = \sum\nolimits_{1 \leq i \leq N,\,\Delta T_i \notin \mathrm{OTI}} C_{c,i} = \sum\nolimits_{1 \leq i \leq N,\,\Delta T_i \notin \mathrm{OTI}} M_{c,i}-D_{c,i}\ln{M_{c,i}},
\end{equation}
where
\begin{equation}
    M_{c,i} = \frac{\Delta t_i}{\Delta T_i} \int_{\Delta T_i} m_c(t) \, \mathrm{d}t,
\end{equation}
$\Delta T_i$ is $i$th time interval, and $\Delta t_i$ is the corresponding live time of the instrument. This fit is performed for each energy channel. After the fit, a $z$-score is introduced to test the merit of the overall light curve model $\sum_c m_c(t)$,
\begin{equation}
    Z_C = \frac{\sum_c \bigl( C_c - C_{c,\textrm{e}} \bigr)}{\sqrt{\sum_c C_{c,\textrm{v}}}} \stackrel{\text {asymp}}{\sim} N(0,1),
\end{equation}
where $C_{c,\textrm{e}}$ and $C_{c,\textrm{v}}$ is the expected mean and variance of $C_c$ \citep[see][]{Kaastra2017CstatTest}. If the light curve model passes the test ($p$-value $\geq$ 0.05), Equation (\ref{Eq1}) is interpolated over the OTI to obtain the estimate of background counts and the corresponding Gaussian uncertainty for each channel. Otherwise, the corresponding occultation data will be discarded in the following analysis. A good fit and a bad fit case are shown in Figure \ref{Fig3}.

\begin{figure}[htb]
    \centering
    \includegraphics[width=0.5\textwidth]{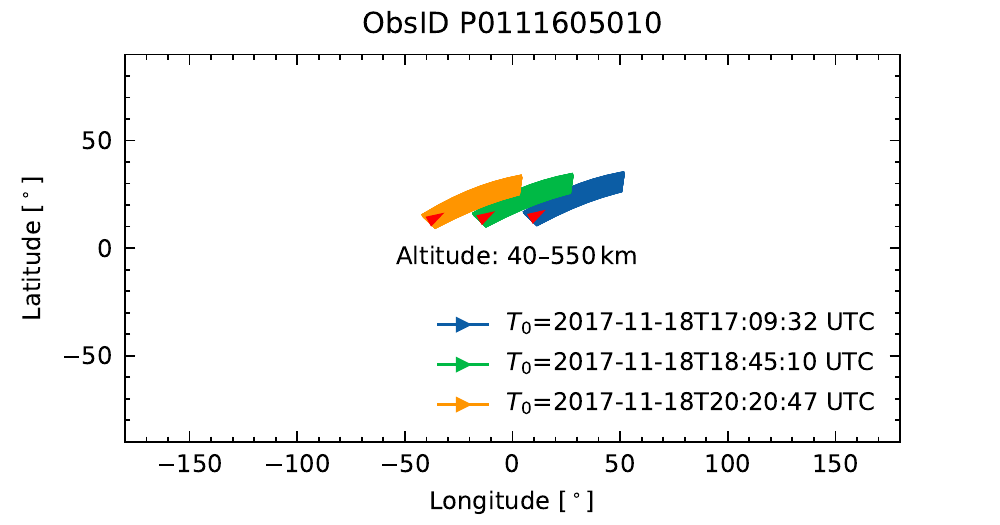}
    \caption{The projections of LoS on the ground are drawn for every 0.5\,s for three occultations in ObsID P0111605010, which make up of the three colored areas in visual. For clarity, only the LoS projections within 40\textendash550\,km altitude are shown. The colored areas indicate the geographical regions whose atmospheric density will be retrieved. The red arrows indicate the viewing direction of telescopes.}
    \label{Fig4} 
\end{figure}

In addition to the cases where the background cannot be well described by Equation (\ref{Eq1}), other data acquired when LE telescope is saturated by bright Earth, or when HE telescope switches to GRB mode, will also be dropped in the following analysis. Furthermore, occultations with the same ObsID are combined to increase photon statistics. In fact, the LoS of occultations with the same ObsID will share a similar latitudinal span. As an example, Figure \ref{Fig4} shows the LoS for three occultations of ObsID P0111605010. Information on all the occultation data analyzed in this paper is summarized in Table \ref{TabA1}, which also details how these data are grouped. The first column represents the ObsID. The second column gives the UTC time when the LoS altitude is 90\,km and the third column depicts the geographical position of the corresponding tangent point. The fourth column provides the latitudinal span of the LoSs, within the altitude range of 40\textendash150\,km. The last two columns offer the available telescopes during the occultation and the corresponding occultation type.

\section{Bayesian Atmospheric Density Retrieval Method} \label{Sect3}
\begin{figure}[htb]
    \centering
    \includegraphics[width=0.5\textwidth]{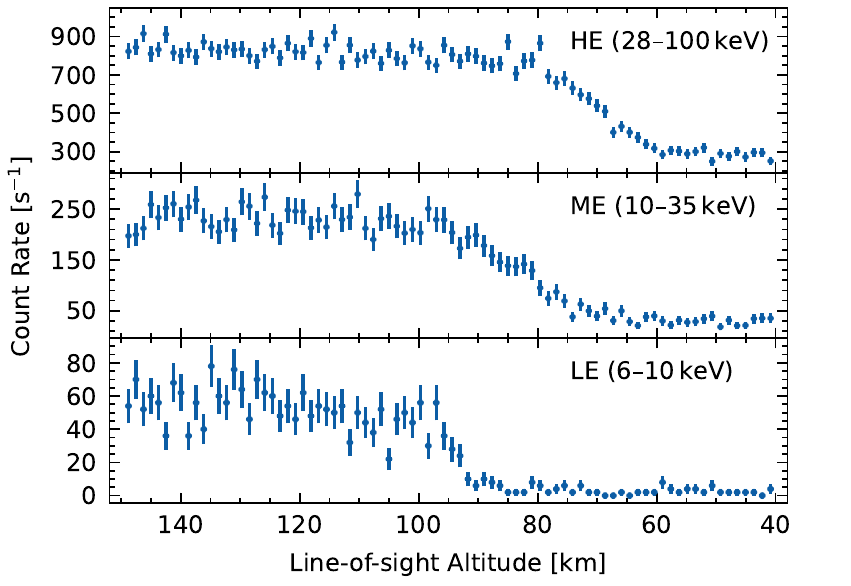}
    \caption{The data point between the dashed and dotted line of Figure \ref{Fig2}. The abscissa axis is transformed from time to corresponding LoS altitude. It should be noted that the X-ray emission received at low LoS altitude is also attenuated by atmosphere layers above the corresponding altitude. For LE, ME, and HE telescopes, the observed X-ray emission is significantly attenuated when the LoS altitude is 90\textendash100\,km, 70\textendash90\,km, and 55\textendash80\,km, respectively.}
    \label{Fig5} 
\end{figure}

When an occultation occurs, X-ray emission with different energies enables ``fluoroscopy'' of atmospheric density at different altitudes. As noticed from Figure \ref{Fig5}, the more energetic the X-ray emission is, the lower the atmosphere through which it can pass. For the occultation of the Crab Nebula, the 6\textendash100\,keV X-ray emission can be considered unattenuated when passing through the atmosphere above 550\,km altitude. In contrast, the emissions detected by LE, ME and HE telescopes are almost extinct when the LoS altitude is below 90\,km, 70\,km, and 55\,km, respectively, leading to low photon statistic and difficulty in retrieving the atmosphere density below these altitudes if data acquired by these three telescopes are independently analyzed. Figure \ref{Fig5} also shows that the attenuation of emission detected by LE, ME, and HE telescopes is not significant when the LoS altitude is above 100\,km, 90\,km, and 80\,km, respectively, which causes similar difficulty in density retrieval above these altitudes. To model atmospheric attenuation, we divide the atmosphere into several layers according to five cases, depending on the data availability of the three telescopes, as listed in Table \ref{Tab1}. The data obtained when the LoS altitude is below the lowest altitude boundary are excluded from next analysis. Therefore, only the densities of the atmosphere between the lowest and highest of the altitude boundaries are considered to be retrieved.

\begin{center}
\begin{table}[htb]
    \centering
    \caption{The five cases of dividing the atmosphere.}
    %\resizebox{0.6\textwidth}{!}{
        \hspace{-8em}\begin{tabular}{|c|ccccccccc|}
        \hline
            Telescope & \multicolumn{9}{c|}{Altitude boundaries [km]}\\
        \hline
            LE+ME+HE & 55 & 65 & 70 & 75 & 80 & 85 & 90 & 100 & 550\\
            LE+ME & & & 70 & 75 & 80 & 85 & 90 & 100 & 550\\
            ME+HE & 55 & 65 & 70 & 75 & 80 & 85 & 90 & & 550\\
            ME &  & & 70 & 75 & 80 & 85 & 90 & & 550\\
            HE & 55 & 65 & 70 & 75 & 80 & & & & 550\\
        \hline
        \end{tabular}
    %}
    \label{Tab1}
\end{table}
\end{center}

The atmospheric attenuation of X-ray photons can be well described by Beer{\textendash}Lambert law
\begin{equation}
    I(E,t) = I_0(E)\,\exp{\left[-\tau(t,E)\right]},
    \label{Eq7}
\end{equation}
where $I(E)$ is the attenuated flux, $I_0(E)$ is the source flux, and $\tau(t,E)$ is the optical depth along the LoS,
\begin{equation}
    \tau(E,t) = \sum_{s} \sigma_{s}(E)\,N_{s}(t),
    \label{Eq8}
\end{equation}
which is the sum of the product between the X-ray cross section $\sigma_{s}(E)$ and the column density $N_{s}(t)$ of each atom species $s$. The column density is defined as
\begin{equation}
    N_{s}(t) = \int_{l_{0}(t)}^{l_{*}}n_{s}(l)\,\mathrm{d}l,
    \label{Eq9}
\end{equation}
where $l_{0}(t)$ is the satellite position at time $t$, $l_{*}$ is the source position, and $n_{s}(l)$ is the number density of atom species $s$ at location $l$ along the LoS.

Equation (\ref{Eq7}), (\ref{Eq8}) and (\ref{Eq9}) indicate that the attenuation is determined by the atmospheric density $n_{s}(l)$ along the LoS. The basis function that we use to fit $n_{s}(l)$ is the NRLMSIS model. Specifically, we first obtain the densities along the LoS from the NRLMSIS model, then the densities are multiplied by a correction factor, that is,
\begin{equation}
    n_{s}(lat_l, lon_l, h_l) = \gamma(h_l)\,n^{\prime}_{s}(lat_l, lon_l, h_l),
    \label{Eq10}
\end{equation}
where $lat_l$, $lon_l$ and $h_l$ are the latitude, longitude and altitude at the location $l$, $n^{\prime}_{s}$ is the density obtained from the NRLMSIS model, and $\boldsymbol{\gamma}$ is an overall density correction factor for all atom species $s$ and is a piecewise function of altitude. The component number $N_\gamma$ of correction factor $\boldsymbol{\gamma}$ is determined by the altitude boundaries listed in Table \ref{Tab1}. For example, if data of LE, ME and HE telescopes are all available for analysis, the correction factor $\boldsymbol{\gamma}$ will have $N_\gamma=8$ components,
\begin{equation}
    \gamma(h) = 
    \begin{cases}
        \gamma_1, & \text{if 55\,km $\leq h < $ 65\,km},\\
        \gamma_2, & \text{if 65\,km $\leq h <$ 70\,km},\\
        \gamma_3, & \text{if 70\,km $\leq h <$ 75\,km},\\
        \gamma_4, & \text{if 75\,km $\leq h <$ 80\,km},\\
        \gamma_5, & \text{if 80\,km $\leq h <$ 85\,km},\\
        \gamma_6, & \text{if 85\,km $\leq h <$ 90\,km},\\
        \gamma_7, & \text{if 90\,km $\leq h <$ 100\,km},\\
        \gamma_8, & \text{if 100\,km $\leq h \leq$ 550\,km}.
    \end{cases}
\end{equation}

\begin{figure}[htb]
    \centering
    \includegraphics[width=0.5\textwidth]{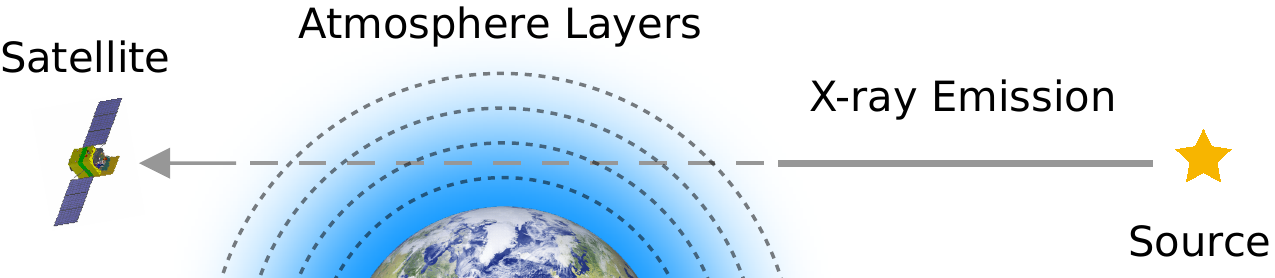}
    \caption{Schematic diagram of X-ray emission passing through several atmosphere layers along the LoS.}
    \label{Fig6} 
\end{figure}

The meaning of Equation (\ref{Eq10}) is that the densities along the LoS obtained from the NRLMSIS model are multiplied by a $\boldsymbol{\gamma}$ factor based on the atmospheric layer to which they belong. Therefore, the left side of Equation (\ref{Eq7}) becomes $I(\boldsymbol{\gamma},E,t)$, and the factor $\boldsymbol{\gamma}$ is added to Equation (\ref{Eq7}), (\ref{Eq8}) and (\ref{Eq9}) accordingly. Figure \ref{Fig6} illustrates the LoS through different layers of atmosphere.

\begin{figure*}[htb]
    \centering
    \setlength\tabcolsep{0pt}
    \begin{tabular}{cc}
        \includegraphics[width=0.5\textwidth]{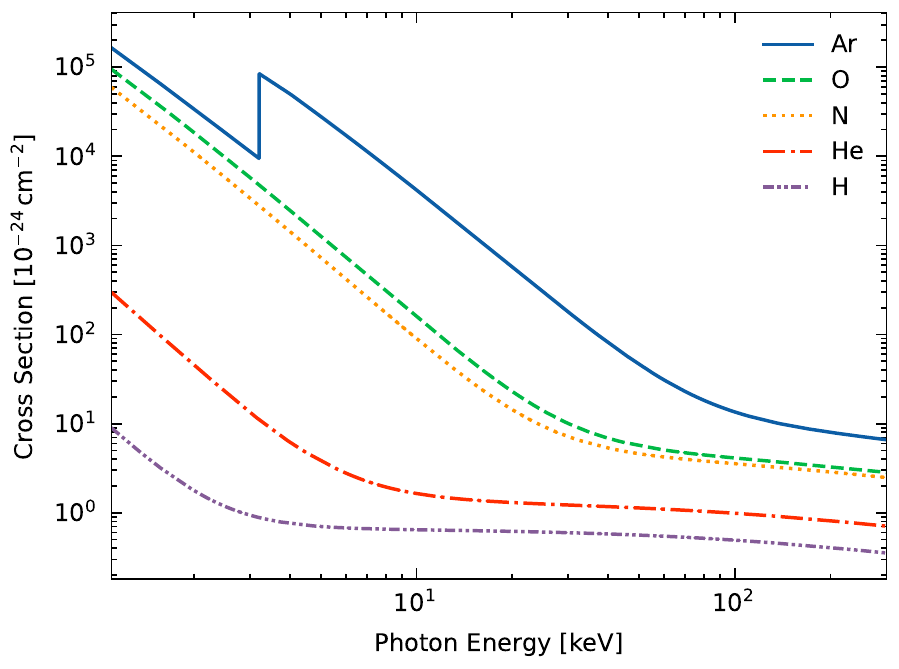}
         & 
        \includegraphics[width=0.5\textwidth]{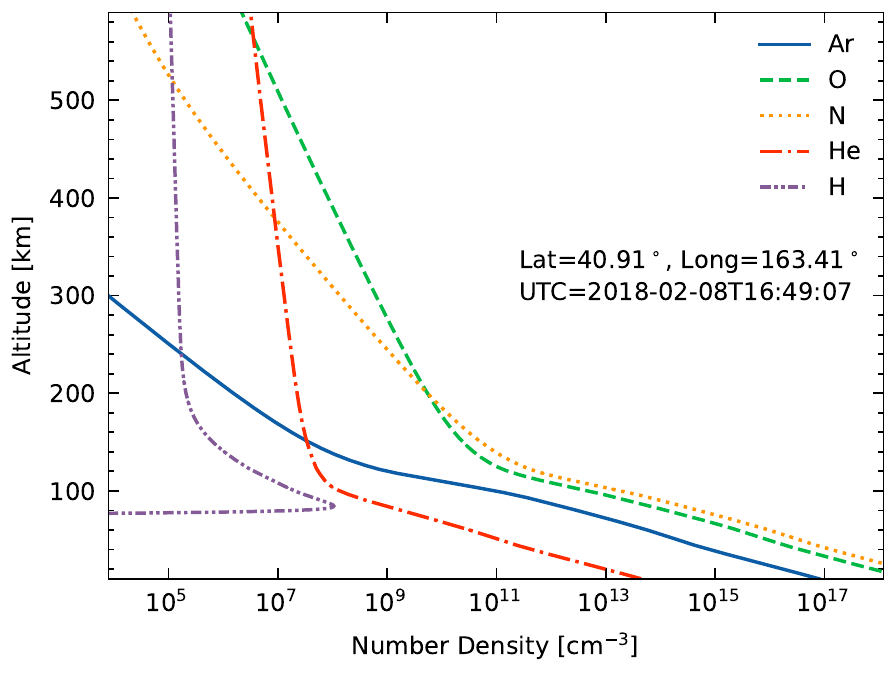}
    \end{tabular}
    \caption{\textit{Left}: The total X-ray cross section of five atoms, including the cross sections of Compton scattering and photo-electric absorption. \textit{Right}: The vertical density profile of five atoms obtained from NRLMSISE-00. Here, the density profile of O includes the contribution from O, O$_2$ and Anomalous O, while the density profile of N includes the contribution from N and N$_2$.}
    \label{Fig7} 
\end{figure*}

In our analysis, the atom species $s$ only include the N, O, and Ar species. The H, He and other species have little contribution to the attenuation of X-ray photons, because of their much lower density and/or smaller X-ray cross section compared to N, O, or Ar. Figure \ref{Fig7} shows the X-ray cross section of H, He, N, O, and Ar, obtained from \texttt{XCOM} database \citep{Berger1987XCOM}, and the corresponding vertical density profile obtained from the NRLMSISE-00 model. The input environmental parameters for the NRLMSIS model, i.e., 10.7\,cm solar radio flux $F10.7$ and geomagnetic indices $Ap$, are obtained from online database \citep{Matzka2021Kp}.

As a consequence, the expected count from the source in channel $c$ during OTI can be modeled as
\begin{equation}
    S_{c,i}(\boldsymbol{\gamma}) = \frac{\Delta t_i}{\Delta Ti}\int_{\Delta {Ti}} \mathrm{d}t \int_E \mathrm{d}E \ R(E,c) \, I(\boldsymbol{\gamma},E,t).
\end{equation}
During $\Delta T_i$, if the background $B_{c,i}$ of channel $c$ is precisely known, the likelihood for observing counts $D_{c,i}$ can be presented by the Poisson distribution as
\begin{equation}
    L\Bigl(D_{c,i}\, \big| \, S_{c,i}(\boldsymbol{\gamma}),B_{c,i} \Bigr) = \frac{\left[S_{c,i}(\boldsymbol{\gamma})+B_{c,i}\right]^{D_{c,i}}}{D_{c,i}!}\exp{\bigl[-\left(S_{c,i}(\boldsymbol{\gamma})+B_{c,i}\right)\bigr]}.
    \label{Eq12}
\end{equation}
However, the exact value of $B_{c,i}$ is unknown and what we learn is usually its estimate. In our case, the background estimate $\hat{B}_{c,i}$ and the corresponding error $\sigma_{c,i}$ can be obtained from the estimation procedure described in Section \ref{Sect2.2}. To reduce the potential statistical bias when inferring the parameter of interest $\boldsymbol{\gamma}$, the uncertainty of the background should be incorporated into Equation (\ref{Eq12}) by taking $B$ as an extra parameter and then being integrated over possible value range. The likelihood of observing counts $D_{c,i}$ is then modified to be an integrated likelihood,
\begin{equation}
    L\left(D_{c,i} \, \middle| \, S_{c,i}(\boldsymbol{\gamma}),\hat{B}_{c,i},\sigma_{c,i} \right) = \int_0^{+\infty} L\left(D_{c,i}\, \middle| \, S_{c,i}(\boldsymbol{\gamma}),B_{c,i} \right) \, L\left(B_{c,i} \, \middle| \, \hat{B}_{c,i},\sigma_{c,i} \right) \, \mathrm{d}B_{c,i},
    \label{Eq13}
\end{equation}
where
\begin{equation}
    L\left(B_{c,i} \, \middle| \, \hat{B}_{c,i},\sigma_{c,i} \right) = \frac{1}{\sqrt{2\pi}\sigma_{c,i}}\exp{\left[-\frac{\left(B_{c,i}-\hat{B}_{c,i}\right)^2}{2\sigma_{c,i}^2}\right]}\,.
\end{equation}
Although Equation (\ref{Eq13}) can be calculated numerically, the time consumption will increase considerably if it is evaluated multiple times. Instead of evaluating the integrated likelihood, we adopt the profile likelihood form of Equation (\ref{Eq13}), that is, we take the maximum of the integrand of Equation (\ref{Eq13}) to be the likelihood. The problem is now reduced to finding a background value $\tilde{B}_{c,i}$ that maximizes the integrand of Equation (\ref{Eq13}) for a given value of $S_{c,i}$. A similar approach is implemented in \texttt{XSPEC} \citep{Arnaud2022XSPEC} to obtain a statistic named \texttt{PGSTAT} to fit Poisson spectral data with Gaussian background. Following the same procedure for deriving the \texttt{PGSTAT} statistic, the profile likelihood derived in our case is
\begin{equation}
    \tilde{L}\left(D_{c,i} \, \middle| \, S_{c,i}(\boldsymbol{\gamma}),\hat{B}_{c,i},\sigma_{c,i} \right) = \frac{\left[S_{c,i}(\boldsymbol{\gamma})+\tilde{B}_{c,i}\right]^{D_{c,i}}}{\sqrt{2\pi}\sigma_{c,i} \, D_{c,i}!}\exp{\left[-\left(S_{c,i}(\boldsymbol{\gamma})+\tilde{B}_{c,i}+\frac{\left(\tilde{B}_{c,i}-\hat{B}_{c,i}\right)^2}{2\sigma_{c,i}^2}\right)\right]}\,,
    \label{Eq15}
\end{equation}
where
\begin{equation}
    \tilde{B}_{c,i} = \frac{1}{2} \Biggl\{\hat{B}_{c,i} - S_{c,i}(\boldsymbol{\gamma}) - \sigma_{c,i}^2 +\biggl\{\Bigl[S_{c,i}(\boldsymbol{\gamma}) + \sigma_{c,i}^2 - \hat{B}_{c,i}\Bigr]^2 - 4\Bigl[\sigma_{c,i}^2 S_{c,i}(\boldsymbol{\gamma}) - \sigma_{c,i}^2 D_{c,i} - \hat{B}_{c,i} S_{c,i}(\boldsymbol{\gamma})\Bigr]\biggr\}^{1/2}\,\Biggr\}\,.
\end{equation}
The profile-out background model $\tilde{B}_{c,i}$ is non-negative when
\begin{equation}
    \hat{B}_{c,i} \big/ \sigma_{c,i}^2 \geq 1-D_{c,i} \big/ S_{c,i}(\boldsymbol{\gamma}),
\end{equation}
which is always met in our analysis, since we have enough data to obtain a background estimate $\hat{B}_{c,i}$ with small uncertainty so that $\hat{B}_{c,i} \big/ \sigma_{c,i}^2 > 1$.

The likelihood of observing the data $\boldsymbol{D}$ during the OTI can be given as
\begin{equation}
    L\left(\boldsymbol{D} \, \middle| \, \boldsymbol{S}(\boldsymbol{\gamma}),\hat{\boldsymbol{B}},\boldsymbol{\sigma} \right) = \prod\nolimits_{1 \leq i \leq N,\,\Delta T_i \in \mathrm{OTI}} \prod\nolimits_{c_{\mathrm{min}} \leq c \leq c_{\mathrm{max}}} \tilde{L}\left(D_{c,i} \, \middle| \, S_{c,i}(\boldsymbol{\gamma}),\hat{B}_{c,i},\sigma_{c,i} \right) \,.
\end{equation}

We adopt Bayesian inference to obtain the posterior distribution of the interested parameter $\boldsymbol{\gamma}$. A flat prior is assumed for $\boldsymbol{\gamma}$, which imposes the constraint that the density should be positive, 
\begin{equation}
    \Theta(\boldsymbol{\gamma}) =
    \begin{cases}
		0, & \text{if any $\gamma_i \leq 0$, for $i=1,\cdots,N_\gamma$}, \\
        1, & \text{otherwise}.
    \end{cases}
\end{equation}
The posterior distribution of $\boldsymbol{\gamma}$ is then
\begin{equation}
    p\left(\boldsymbol{\gamma} \, \middle| \, \boldsymbol{D},\boldsymbol{S}(\boldsymbol{\gamma}),\hat{\boldsymbol{B}},\boldsymbol{\sigma} \right) =\frac{L\left(\boldsymbol{D} \, \middle| \, \boldsymbol{S}(\boldsymbol{\gamma}),\hat{\boldsymbol{B}},\boldsymbol{\sigma} \right) \, \Theta(\boldsymbol{\gamma})}{\int_{\mathbb{R}^{N_\gamma}} L\left(\boldsymbol{S}(\boldsymbol{\gamma}) \, \middle| \, \boldsymbol{D},\hat{\boldsymbol{B}},\boldsymbol{\sigma} \right) \, \Theta(\boldsymbol{\gamma}) \,\mathrm{d}\boldsymbol{\gamma}} \propto L\left(\boldsymbol{D} \, \middle| \, \boldsymbol{S}(\boldsymbol{\gamma}),\hat{\boldsymbol{B}},\boldsymbol{\sigma} \right) \, \Theta(\boldsymbol{\gamma}) \, .
\end{equation}
Note that if multiple occultation data are available for analysis, then these data can be incorporated together into the likelihood term of the posterior. We adopt the Markov chain Monte Carlo (MCMC) technique to sample the posterior, since there is no analytic way to evaluate the high-dimensional posterior. A Python package \texttt{emcee} \citep{Foreman-Mackey2013EmceeMCMC} is used to draw samples from the posterior. In our analysis, we run a total of 40 ensemble samplers of \texttt{emcee} for 20000 iterations, and discard the first 2000 iterations in the posterior analysis, which are taken as burn-in sampling.

\begin{center}
\begin{figure*}
    \centering
    \includegraphics[width=1\textwidth]{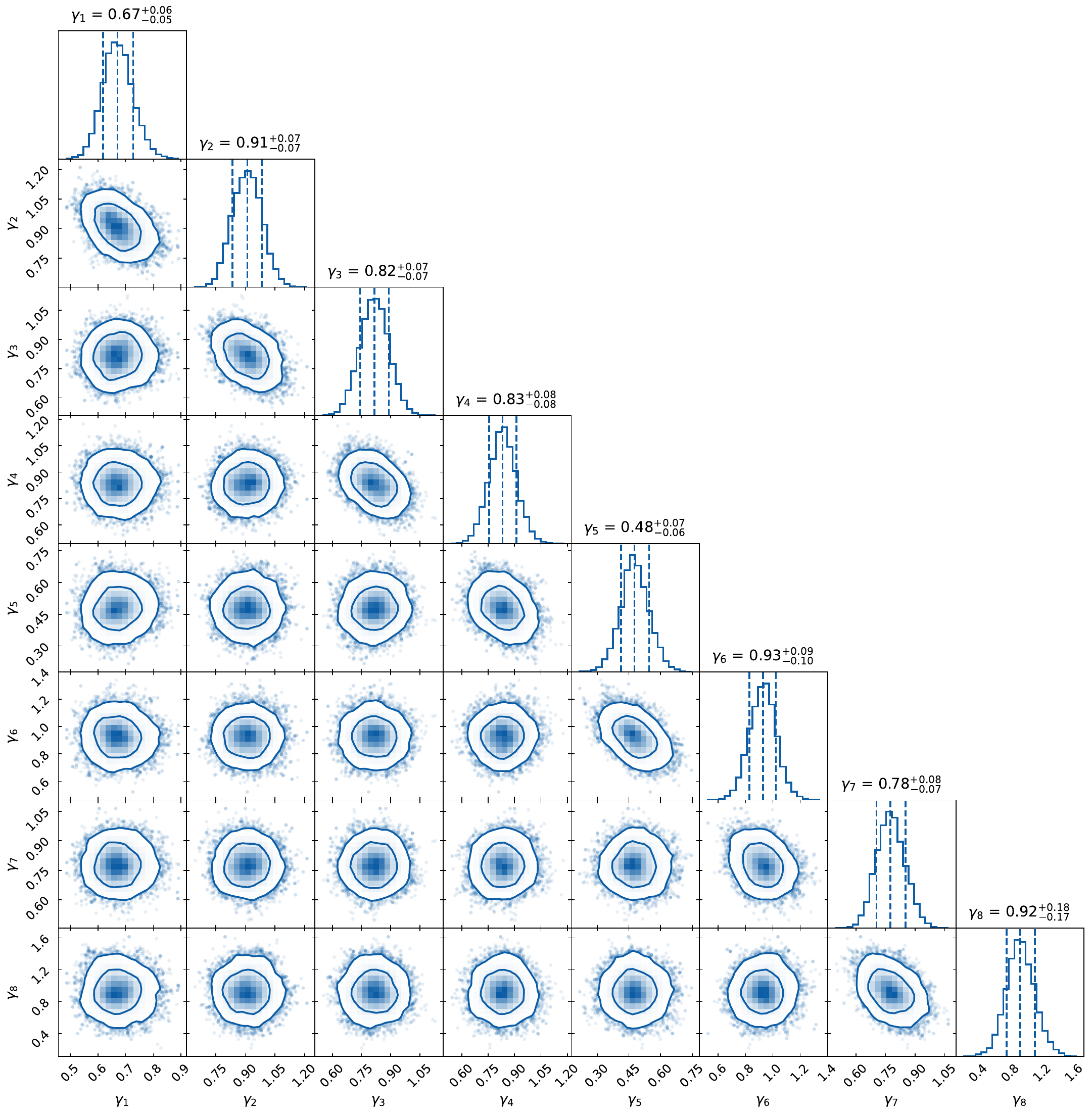}
    \caption{The marginal distribution of density correction factors $\boldsymbol{\gamma}$. The dashed lines in the 1-D histogram are the 15.8\%, 50\% and 84.1\% quantiles of each parameter, which are also given at the top of the histogram. On the 2-D plots, the two contours indicate the 68.3\% and 95.0\% credible regions. The density correction coefficients for adjacent atmosphere layers exhibit the negative correlation as expected.} %The plot is made with \texttt{corner.py} \citep{Foreman-Mackey2016CornerPy}.}
    \label{Fig8}
\end{figure*}
\end{center}

\begin{center}
\begin{figure}[htb]
    \centering
    \includegraphics[width=0.5\textwidth]{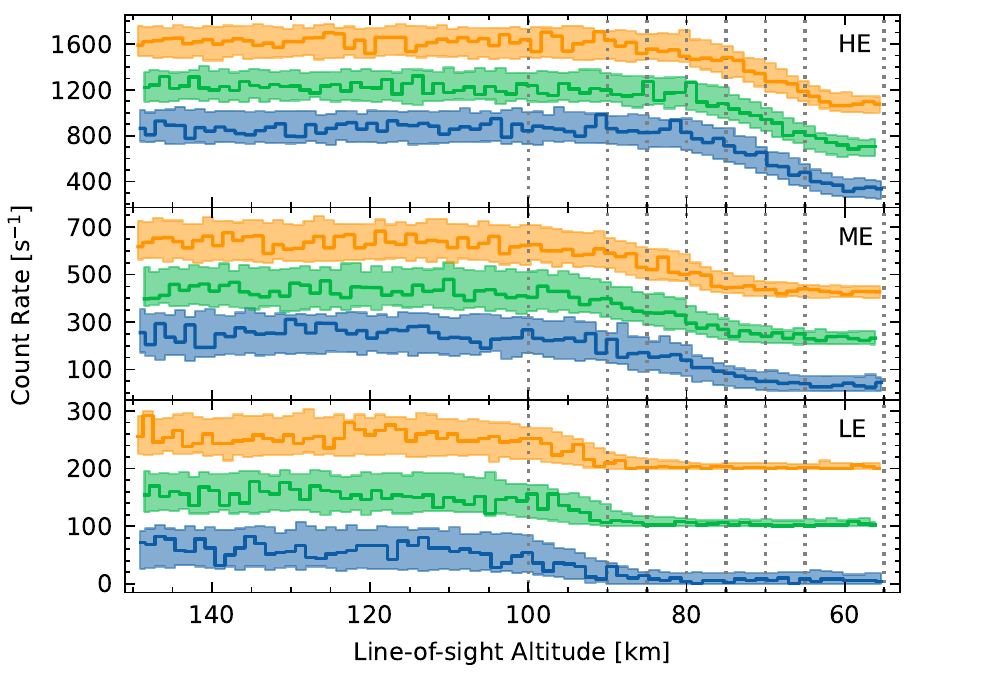}
    \caption{Observed light curve data (solid lines) and corresponding posterior predictive light curves ranges (shaded area). The blue, green and orange data are obtained from the first, second and third occultations of observation P0111605046. The shaded regions are obtained from 1000 posterior predictive simulations. For clarity, the data point and the corresponding predictive light curves are shifted up by 100 (LE), 200 (ME), or 400 (HE) for the second occultation, and are shifted up by 200 (LE), 400 (ME), or 800 (HE) for the third occultation.}
    \label{Fig9}
\end{figure}
\end{center}

\begin{center}
\begin{figure*}[htb]
    \centering
    \includegraphics[width=0.8\textwidth]{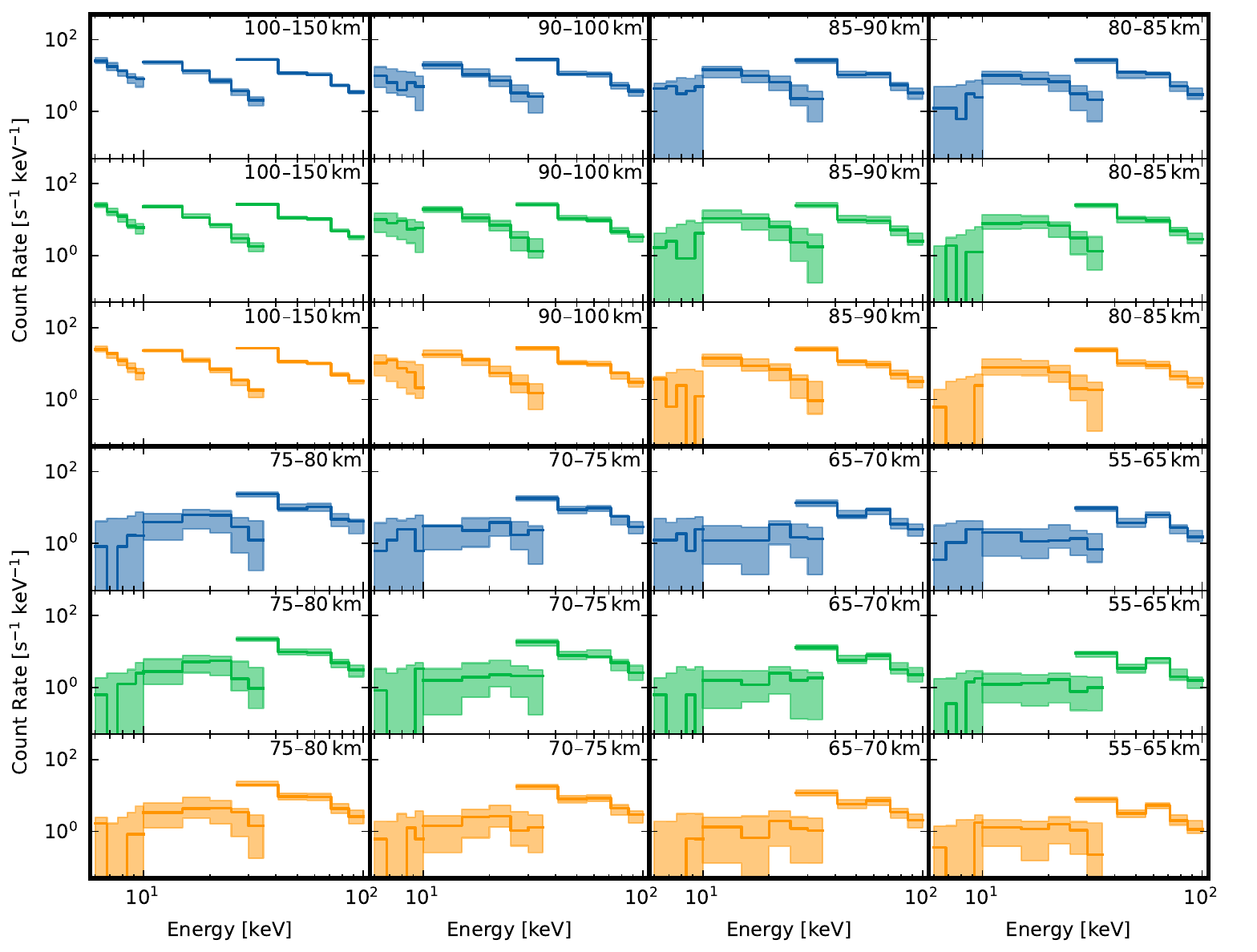}
    \caption{Observed spectral data (solid lines) and corresponding posterior predictive spectra ranges (shaded area). The blue, green and orange data are obtained from the first, second and third occultations of observation P0111605046. These spectra are generated by grouping data according to the LoS altitude, which is annotated in the upper right of each subplot. For clarity, the spectral data are binned into 5 points for each telescope. The shaded regions are obtained from the same 1000 simulations as in Figure \ref{Fig9}.}
    \label{Fig10}
\end{figure*}
\end{center}

As an example, we model the three occultations taken place in observation P0111605046, and obtain the posterior samples of $\boldsymbol{\gamma}$, as shown in Figure \ref{Fig8}. The negative correlation of density correction coefficients for adjacent atmosphere layers is clearly seen as expected. Previous work used light-curve or spectral fitting method that utilized only partial information from the data and did not consider this correlation. Since the atmospheric density of 100\textendash550\,km is not directly interested, but this atmosphere layer still has some contribution to the absorption of X-ray photons, the corresponding density correction factor, $\gamma_8$, is considered a nuisance parameter and marginalized when inferring the density correction factors of low atmosphere layers. This could help to accurately retrieve the density of the low-atmosphere layers of interest. Our method can be considered as either a joint light curve fitting with a high spectral resolution, or a joint spectral fitting with a high temporal resolution, to simultaneously retrieve the density of different atmosphere layers, which makes full use of data, including the spectral and temporal information. Figure \ref{Fig9} and Figure \ref{Fig10} show the light curves and spectra predicted from our Bayesian models, which are generally consistent with the observed data.

\section{Results and Discussion} \label{Sect4}

\begin{figure*}[htb]
    \centering
    \setlength\tabcolsep{0pt}
    \begin{tabular}{cc}
        \includegraphics[width=0.5\textwidth]{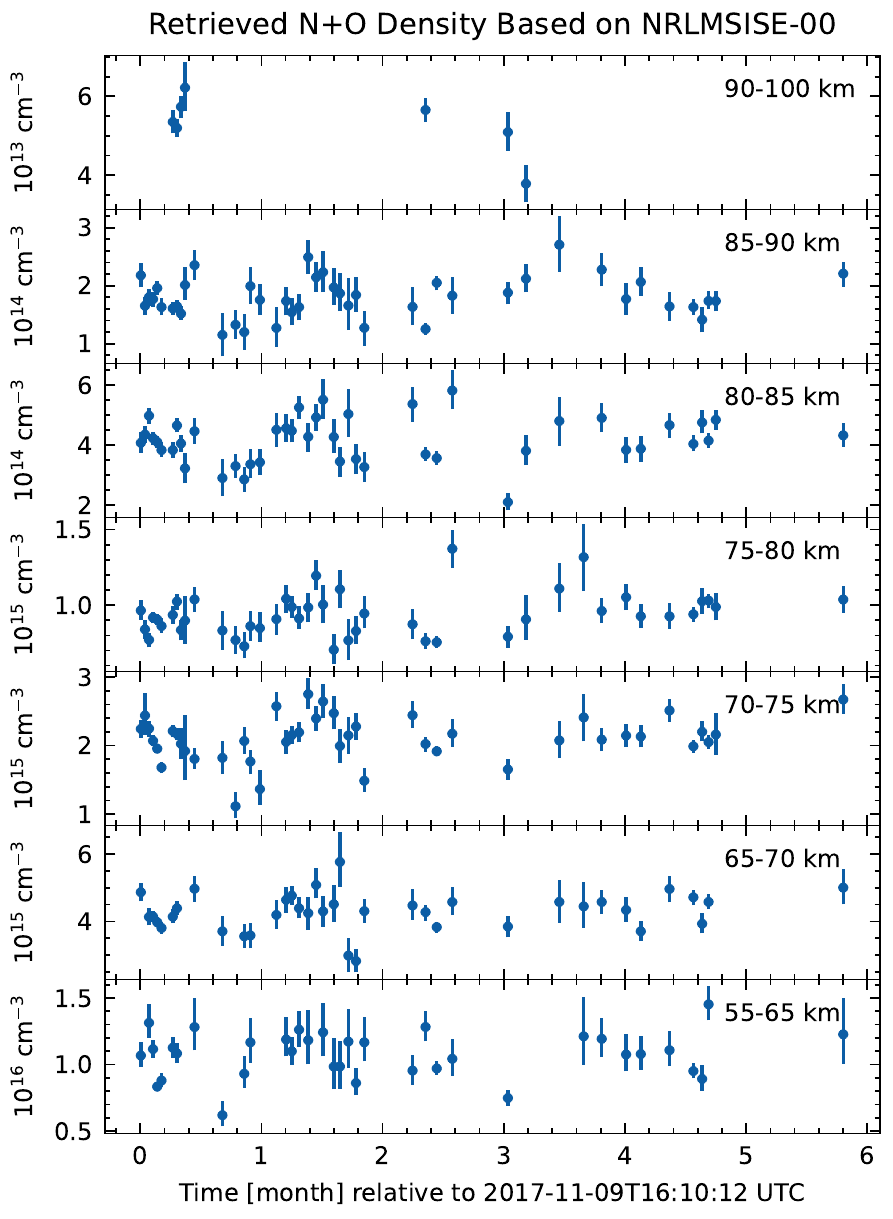}
         & 
        \includegraphics[width=0.5\textwidth]{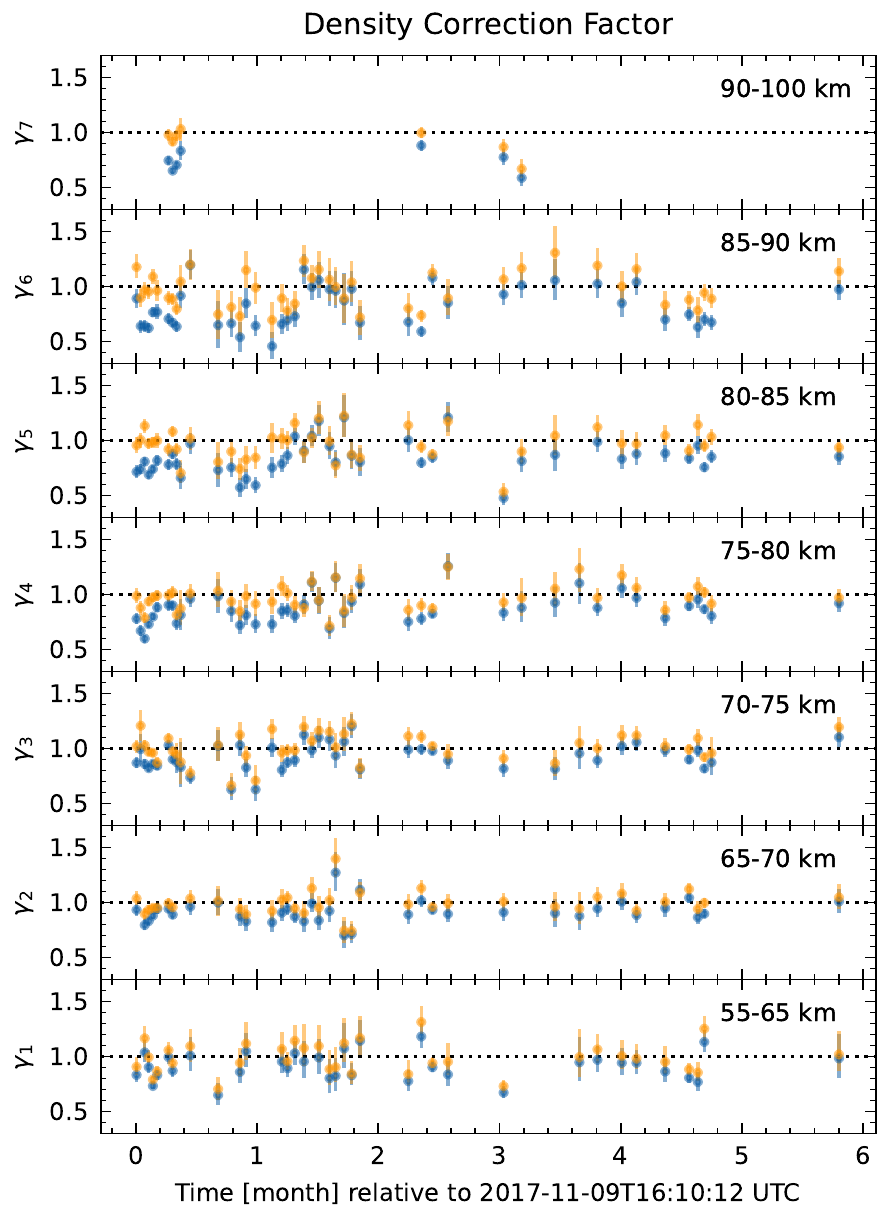}
    \end{tabular}
    \caption{\textit{Left}: The retrieved densities based on NRLMSISE-00 models. Since the density of Ar is less than 1\% when compared to N+O, only N+O density is shown here. For clarity, the retrieved densities of observation P0111605063 (the last data points) are shifted left by 4 months. \textit{Right}: The density correction factors of NRLMSISE-00 model (in blue) and NRLMSIS 2.0 model (in orange). For clarity, the correction factors obtained from observation P0111605063 (the last data points) are shifted left by 4 months.}
    \label{Fig11}
\end{figure*}

Using the density retrieval method described in Section \ref{Sect3}, we analyzed a total of 115 occultations that occurred during 45 observations listed in Table \ref{TabA1}, and obtain density measurements in the altitude range of 55\textendash100\,km and the latitude of $-20^\circ$\textendash$60^\circ$. The left panel of Figure \ref{Fig11} displays all the retrieved neutral densities (N+O) of each atmosphere layer as a function of time. It should be noted that if there is a significant density structure across a layer, the retrieved density presented here is just the averaged approximation. The retrieved densities are found to be systematically lower than the NRLMSISE-00 model. The densities that we retrieved are $\sim$90\%, $\sim$80\% and $\sim$75\% of the NRLMSISE-00 model in the altitude range of 55\textendash80\,km, 80\textendash90\,km and 90\textendash100\,km, respectively, as shown in the right panel of Figure \ref{Fig11}. The retrieved neutral densities are generally consistent with the NRLMSIS 2.0 model in the altitude range of 55\textendash90\,km. Due to the fact that retrieval performed within 90\textendash100\,km is relatively scarce, it is difficult to draw a conclusion whether the densities predicted from the NRLMSIS 2.0 model are generally correct.

\cite{Determan2007AtmosphericDensity} analyzed the 17.9\textendash42.6\,keV data of an occultation of the Crab Nebula observed by RXTE/PCA in November 2005, and their light curve fitting method yielded a density correction factor of 0.86$\pm$0.12 for NRLMSISE-00 in the altitude range of 70\textendash90\,km. Another correction of NRLMSISE-00 based on the light curve fitting method gives a factor of 0.88$\pm$0.07 in the altitude range of 85\textendash100\,km \citep{Yu2022AtmosphericDensity2}, using the 6\textendash10\,keV data acquired with \textit{Insight}-HXMT/LE during an occultation of the Crab Nebula in September 2018. \cite{Yu2022AtmosphericDensity1} conducted a direct density retrieval by the spectral fitting method using the data observed by \textit{Insight}-HXMT/LE during an occultation of the Crab Nebula in November 2017, and obtained a correction factor of 0.86$\pm$0.03 for NRLMSISE-00 in the altitude range of 90\textendash100\,km. \cite{Katsuda2021AtmosphericDensity} applied an inversion method to the column densities obtained from the 219 occultation spectra acquired with Suzaku/XIS and Hitomi/HXI, and found that the inverted number density profiles are $\sim$50\% smaller than NRLMSISE-00 in the altitude range of 70\textendash110\,km.

 The previous results coincide with ours in some altitude ranges and times, although three of them, i.e. \cite{Determan2007AtmosphericDensity}, \cite{Yu2022AtmosphericDensity1} and \cite{Yu2022AtmosphericDensity2}, contain only an independent measurement of a single atmosphere layer. However, our results do not support the previous claim of the $\sim$50\% density deficit of the NRLMSISE-00 model \citep{Katsuda2021AtmosphericDensity}. The retrieved densities are found to be approximately half of the NRLMSISE-00 model in only a minor fraction of our observations. This finding, although preliminary, suggests that the density deficit in the altitude range of 70\textendash100\,km could reach $\sim$50\% but untypical. In addition, it can be seen from the right panel of Figure \ref{Fig11} that the disparity between the retrieved neutral densities and the NRLMSISE-00 model increases with altitude. This trend is similar to the results of 70\textendash100\,km in \cite{Katsuda2021AtmosphericDensity}, while the correction factors for the NRLMSIS 2.0 model do not exhibit this trend. In addition, the predicted values from the NRLMSISE-00 model are sometimes close to the retrieved densities, e.g. the data points around 4 months relative to zero time point; but sometimes not, e.g. those around 1 month relative to zero time point, this may be attributed to the imperfect modeling of seasonal variations. The NRLMSIS 2.0 model seems to have fixed this issue.

Although the previous results are consistent with ours to some extent, they may be somewhat limited by potential systematic uncertainty. As pointed out by \cite{Katsuda2021AtmosphericDensity}, using a single correction factor at all altitudes when fitting the occultation light curve would result in some systematic uncertainties. In addition, light curves of different energy bands may sample overlapping atmosphere layers, which means that the retrieved densities would require some regularization, as discussed in \cite{Determan2007AtmosphericDensity}. Regarding the spectral fitting method, the direct retrieval method \citep{Yu2022AtmosphericDensity1} may also be subject to systematic uncertainty without considering tomography. \cite{Katsuda2021AtmosphericDensity} adopted the inversion method \citep{Roble1972TechniqueRecovering} in their occultation spectral fitting to obtain the density profile, but this technique is, in fact, not a full tomography, i.e. the spectra obtained at lower LoS altitude do not contribute to the retrieval of the density of the higher atmosphere layers. In fact, the lower-energy part of the spectra obtained in a lower LoS altitude can also be used to refine the density measurements of higher altitude, since the higher atmosphere layers will absorb some low-energy emission.

The density retrieval method we developed solves the issues discussed above. The absorption of the X-ray emission by multiple atmosphere layers is modeled together, and the corresponding densities are retrieved simultaneously in our method. This effectively reduces the systematic bias that can arise from applying only one single correction factor to density profile over a wide altitude range, which is the case that can be encountered with the light curve fitting method used in previous work, e.g. \cite{Determan2007AtmosphericDensity} and \cite{Yu2022AtmosphericDensity2}. We also fundamentally avoid the regularization issue of the light curve fitting method when the light curve data with different energy sampled overlapping altitudes. We also note that the attenuation effect is cumulative across the atmosphere layers along the LoS, meaning that data from each time bin and energy channel can contribute more or less to the constraints on all the density correction factors. In other words, the data obtained at higher LoS altitude are primarily used to determine the density of the corresponding layer, but also further affect the parameter ranges of the lower atmosphere layers, while the data obtained at lower LoS altitude indeed can help to better constrain the densities of the higher atmosphere layers. It should be noted that the inversion method adopted in \cite{Katsuda2021AtmosphericDensity} only considers the former point, while both the former and latter points are well handled in this paper. Moreover, as already demonstrated in Section \ref{Sect3}, our method is more generic than the light curve or spectral fitting method, which can make almost full use of the spectral and temporal information of the occultation data, and therefore the statistical uncertainties of the parameters of interest would be smaller.

The density retrieval method we developed is also generic in terms of atmospheric model selection. As shown in the right panel of Figure \ref{Fig7}, the density profile of atmospheric species decreases almost exponentially with altitude. Therefore, the density profile of a single atmosphere layer $k$ can be modeled as
\begin{equation}
    n_s = a_{s,k} \exp{(-b_{s,k}\,h)},
    \label{Eq22}
\end{equation}
where $a_{s,k}$ and $b_{s,k}$ are the parameters of the density profile of the species $s$ at the atmosphere layer $k$, and $h$ is the altitude. This is the function used to invert the density profile from the spectral fit in \cite{Katsuda2021AtmosphericDensity}. Considering that the density profile should be continuous between atmospheric layers and the relative abundance ratios of N, O and Ar could be fixed to the predicted value of the NRLMSISE model (or take the relative abundance ratios as free parameters and then marginalize them out in the posterior distribution), the free parameters to model the neutral densities of $k$ atmosphere layers are reduced to a normalization factor $a$ and decreasing indices $b_k$ with number $k$. A similar density retrieval process can be performed after replacing the term $n_s$ in Equation (\ref{Eq9}) with Equation (\ref{Eq22}), providing another cross-check for the NRLMSIS model.

Furthermore, it is possible to apply our density retrieval method to the occultation data acquired by all-sky FoV satellites, as long as the background can be well fitted and the spectral model of the target source is known. The spectral model can be obtained from the spectral analysis of the target source, using pointed observations from other satellites at the same time, or simply by assuming a reasonable source model from previous knowledge, where the parameters of the source model are additional parameters used to describe the occultation process. Although the latter case will obviously lead to degeneracy between the source and density parameters, it seems to be the only option for the EOT density retrieval using occultation data from all-sky FoV satellites when precise knowledge of the source spectra is unavailable.

\section{Summary} \label{Sect5}

In this paper, we proposed a Bayesian density retrieval method based on the EOT, and then applied the method to \textit{Insight}-HXMT observations of Earth occulation of the Crab Nebula and obtained the neutral atmospheric density measurement within 55\textendash100\,km. The neutral densities of different altitude layers are simultaneously measured by modeling the photon attenuation during the occultation process. Unlike previous work in which the densities of different altitude layers were measured independently, our joint fitting method reduces potential systematic bias. Furthermore, the method we developed has the advantages of making full use of the data information and the capability to be applied to all-sky FoV satellites.

We find that the neutral density retrieved is typically about 10\%, 20\%, and 25\% less than the values of the NRLMSISE-00 model in the altitude range of 55–80 km, 80-90 km and 90–100 km, respectively. The density deficit of the NRLMSISE-00 model can be up to $\sim$50\% in some regions. We also note that the densities retrieved are generally consistent with the newly released NRLMSIS 2.0 model \citep{Emmert2021NRLMSIS}.

Our study does not include the 1\textendash6\,keV occultation data of \textit{Insight}-HXMT/LE telescope, which can be used to retrieve neutral densities above 100\,km. This is largely due to the fact that the variation of cosmic diffuse X-ray background during the occultation cannot be well predicted by a simple continuum. Since the atmosphere above 150\,km altitude is dependent on the solar cycle \citep[e.g.,][]{Meier2015RemoteSensing}, the future work will focus on developing an effective background estimate method of 1\textendash6\,keV data of LE telescope during the occultation, so that these data can be used to test if the NRLMSIS model accurately reflects the influence of the solar cycle on the atmosphere above 150\,km.

All-sky FoV satellites, e.g. GECAM \citep{li2022technology} and \textit{Fermi}/GBM \citep{Meegan2009GBM}, have accumulated a large amount of occultation data that can be used to study the long-term evolution of the Earth atmosphere. It is also necessary to extend the density retrieval method to apply to all-sky FoV satellites, and this is a subject of ongoing research.

\section*{Acknowledgements}

This work used data from the \textit{Insight}-HXMT mission, a project funded by the China National Space Administration (CNSA) and the Chinese Academy of Sciences (CAS).  We gratefully acknowledge the support from the National Program on Key Research and Development Project (Grant No.2021YFA0718500) from the Minister of Science and Technology of China (MOST). The authors thank supports from the National Natural Science Foundation of China under Grants 12273043, U1838201, U1838202, U1938109, U1938102, U1938108, U1938111, 12103055 and 41604152. This research has been supported by the Youth Innovation Promotion Association of the Chinese Academy of Sciences (grant no. 2018178).

\clearpage
\newpage
\bibliography{reference}

\appendix
\restartappendixnumbering
\section{Occultation Data List}
\begin{center}
\begin{longtable}{cccccc}
\caption{Summary of occultation data analyzed in this paper.}\label{TabA1}\\
\hline
ObsID & UTC$^a$ & \begin{tabular}{c}Tangent point$^b$ \\ Lat, Long [$^\circ$] \end{tabular} & Lat span$^c$ [$^\circ$] & Telescope & Type$^d$ \\
\hline
\endfirsthead
\hline
ObsID & UTC$^a$ & \begin{tabular}{c}Tangent point$^b$ \\ Lat,\,Long [$^\circ$] \end{tabular} & Lat span$^c$ [$^\circ$] & Telescope & Type$^d$ \\
\hline
\endhead
\hline
\multicolumn{6}{l}{\small $^a$UTC time when LoS altitude $h = 90$\,km.}\\
\multicolumn{6}{l}{\small $^b$Latitude and longitude of the tangent point at $h = 90$\,km.}\\
\multicolumn{6}{l}{\small $^c$Latitudinal span of LoS, within altitude 40\textendash150\,km.}\\
\multicolumn{6}{l}{\small $^d$The letter S stands for Setting and the letter R stands for Rising.}\\
\endfoot
\hline
\multicolumn{6}{l}{\small $^a$UTC time when LoS altitude $h = 90$\,km.}\\
\multicolumn{6}{l}{\small $^b$Latitude and longitude of the tangent point at $h = 90$\,km.}\\
\multicolumn{6}{l}{\small $^c$Latitudinal span of LoS, within altitude 40\textendash150\,km.}\\
\multicolumn{6}{l}{\small $^d$The letter S stands for Setting and the letter R stands for Rising.}\\
\endlastfoot
P0111605001 & 2017-11-09T16:10:12 & -9.85,\,-121.69 & (-18.81,\,0.06) & HE & S \\
 & 2017-11-09T19:21:11 & -8.98,\,-169.23 & (-18.03,\,0.90) & ME+HE & S \\
 & 2017-11-09T22:32:15 & -8.13,\,143.23 & (-17.24,\,1.71) & ME+HE & S \\
 & 2017-11-10T00:07:48 & -7.69,\,119.46 & (-16.85,\,2.12) & ME & S \\
\hline
P0111605002 & 2017-11-10T17:38:58 & -2.47,\,-141.90 & (-12.10,\,7.14) & ME & S \\
 & 2017-11-10T19:14:32 & -1.97,\,-165.65 & (-11.63,\,7.63) & ME & S \\
 & 2017-11-10T22:25:32 & -0.94,\,146.85 & (-10.70,\,8.63) & ME & S \\
 & 2017-11-11T00:01:06 & -0.41,\,123.11 & (-10.23,\,9.13) & ME & S \\
\hline
P0111605003 & 2017-11-11T17:32:35 & 5.81,\,-137.92 & (-4.65,\,15.25) & ME & S \\
 & 2017-11-11T19:08:10 & 6.43,\,-161.63 & (-4.10,\,15.83) & ME & S \\
 & 2017-11-11T22:19:13 & 7.66,\,150.95 & (-3.03,\,17.05) & ME+HE & S \\
 & 2017-11-11T23:54:48 & 8.27,\,127.24 & (-2.48,\,17.64) & ME+HE & S \\
\hline
P0111605004 & 2017-11-12T17:26:36 & 15.51,\,-133.32 & (3.87,\,24.91) & ME+HE & S \\
 & 2017-11-12T19:02:13 & 16.21,\,-156.98 & (4.48,\,25.62) & ME+HE & S \\
 & 2017-11-12T22:13:18 & 17.57,\,155.69 & (5.68,\,27.02) & ME+HE & S \\
 & 2017-11-12T23:48:55 & 18.27,\,132.03 & (6.29,\,27.73) & ME+HE & S \\
\hline
P0111605005 & 2017-11-13T18:56:38 & 27.02,\,-151.41 & (13.77,\,36.74) & ME+HE & S \\
 & 2017-11-13T20:32:16 & 27.76,\,-174.99 & (14.41,\,37.51) & ME+HE & S \\
 & 2017-11-13T22:07:42 & 28.50,\,161.43 & (15.02,\,38.28) & ME+HE & S \\
 & 2017-11-13T23:43:20 & 29.23,\,137.85 & (15.64,\,39.05) & ME+HE & S \\
\hline
P0111605006 & 2017-11-14T22:02:27 & 39.40,\,168.53 & (24.07,\,49.93) & ME+HE & S \\
 & 2017-11-14T23:38:04 & 40.07,\,145.05 & (24.65,\,50.67) & ME+HE & S \\
\hline
P0111605009 & 2017-11-17T17:15:02 & 35.09,\,21.70 & (20.49,\,45.30) & LE+ME+HE & R \\
 & 2017-11-17T18:50:40 & 34.37,\,-1.82 & (19.89,\,44.53) & LE+ME+HE & R \\
\hline
P0111605010 & 2017-11-18T17:09:32 & 24.04,\,28.21 & (11.19,\,33.64) & LE+ME+HE & R \\
 & 2017-11-18T18:45:10 & 23.29,\,4.60 & (10.57,\,32.89) & LE+ME+HE & R \\
 & 2017-11-18T20:20:47 & 22.57,\,-19.02 & (9.94,\,32.15) & LE+ME+HE & R \\
\hline
P0111605011 & 2017-11-19T17:03:40 & 13.43,\,33.50 & (2.03,\,22.83) & LE+ME & R \\
 & 2017-11-19T18:39:17 & 12.76,\,9.82 & (1.44,\,22.17) & LE+ME & R \\
 & 2017-11-19T20:14:53 & 12.11,\,-13.87 & (0.87,\,21.49) & LE+ME & R \\
\hline
P0111605012 & 2017-11-20T18:33:11 & 3.44,\,14.23 & (-6.82,\,12.90) & LE+ME & R \\
\hline
P0111605014 & 2017-11-23T02:17:26 & -13.03,\,-97.36 & (-21.76,\,-2.92) & ME+HE & R \\
 & 2017-11-23T03:52:58 & -13.38,\,-121.15 & (-22.12,\,-3.26) & ME+HE & R \\
\hline
P0111605018 & 2017-11-30T00:47:16 & 51.54,\,123.61 & (36.26,\,60.44) & ME+HE & S \\
\hline
P0111605020 & 2017-12-03T05:07:45 & 45.20,\,48.78 & (31.44,\,53.16) & ME & S \\
 & 2017-12-03T06:43:11 & 45.05,\,24.73 & (31.32,\,52.99) & ME & S \\
 & 2017-12-03T08:18:37 & 44.90,\,0.67 & (31.20,\,52.83) & ME & S \\
\hline
P0111605021 & 2017-12-05T09:36:59 & 39.93,\,-24.93 & (27.22,\,47.43) & ME+HE & S \\
 & 2017-12-05T11:12:26 & 39.76,\,-48.98 & (27.08,\,47.25) & ME+HE & S \\
\hline
P0111605022 & 2017-12-06T22:12:02 & 35.80,\,141.98 & (23.80,\,43.06) & ME+HE & S \\
 & 2017-12-06T23:47:28 & 35.61,\,117.94 & (23.66,\,42.86) & ME+HE & S \\
\hline
P0111605023 & 2017-12-09T05:52:30 & 28.74,\,20.43 & (17.79,\,35.79) & ME & S \\
 & 2017-12-09T07:27:56 & 28.52,\,-3.61 & (17.59,\,35.57) & ME & S \\
 & 2017-12-09T09:03:23 & 28.31,\,-27.65 & (17.42,\,35.35) & ME & S \\
\hline
P0111605025 & 2017-12-13T08:30:33 & 14.11,\,-30.12 & (4.75,\,21.26) & ME+HE & S \\
 & 2017-12-13T10:06:01 & 13.86,\,-54.16 & (4.52,\,21.02) & ME+HE & S \\
\hline
P0111605026 & 2017-12-15T17:47:10 & 4.76,\,-175.53 & (-3.89,\,12.30) & ME+HE & S \\
 & 2017-12-15T19:22:38 & 4.49,\,160.43 & (-4.14,\,12.05) & ME+HE & S \\
 & 2017-12-15T20:58:06 & 4.24,\,136.40 & (-4.37,\,11.80) & ME+HE & S \\
\hline
P0111605027 & 2017-12-17T03:11:56 & -0.75,\,39.68 & (-9.05,\,7.09) & ME+HE & S \\
 & 2017-12-17T04:47:24 & -1.02,\,15.64 & (-9.30,\,6.84) & ME+HE & S \\
 & 2017-12-17T06:22:51 & -1.27,\,-8.39 & (-9.53,\,6.60) & HE & S \\
\cline{2-6}
 & 2017-12-17T03:47:52 & -14.34,\,-143.10 & (-22.00,\,-5.44) & ME+HE & R \\
 & 2017-12-17T05:23:20 & -14.11,\,-167.13 & (-21.77,\,-5.24) & ME+HE & R \\
\hline
P0111605029 & 2017-12-18T22:09:29 & -7.64,\,110.72 & (-15.58,\,0.65) & ME+HE & S \\
 & 2017-12-19T00:20:51 & -7.71,\,-95.97 & (-15.64,\,0.59) & ME+HE & R \\
 & 2017-12-19T01:20:25 & -8.13,\,62.65 & (-16.04,\,0.19) & ME+HE & S \\
 & 2017-12-19T01:56:19 & -7.46,\,-120.00 & (-15.40,\,0.81) & ME+HE & R \\
\hline
P0111605030 & 2017-12-21T04:51:11 & 0.74,\,-169.15 & (-7.64,\,8.50) & ME & R \\
 & 2017-12-21T06:26:39 & 1.00,\,166.82 & (-7.39,\,8.75) & ME+HE & R \\
 & 2017-12-21T08:02:07 & 1.27,\,142.78 & (-7.15,\,8.99) & ME+HE & R \\
\hline
P0111605031 & 2017-12-23T02:59:38 & 8.35,\,-146.26 & (-0.55,\,15.74) & ME+HE & R \\
 & 2017-12-23T04:35:05 & 8.61,\,-170.29 & (-0.30,\,15.99) & ME & R \\
 & 2017-12-23T06:10:33 & 8.88,\,165.66 & (-0.07,\,16.23) & ME+HE & R \\
 & 2017-12-23T07:46:01 & 9.13,\,141.63 & (0.18,\,16.48) & ME+HE & R \\
\hline
P0111605032 & 2017-12-24T21:56:59 & 15.30,\,-75.32 & (5.83,\,22.41) & ME+HE & R \\
 & 2017-12-24T23:32:27 & 15.55,\,-99.36 & (6.06,\,22.66) & ME+HE & R \\
\hline
P0111605033 & 2017-12-27T13:35:09 & 24.90,\,43.07 & (14.44,\,31.92) & ME+HE & R \\
 & 2017-12-27T15:10:36 & 25.13,\,19.02 & (14.64,\,32.14) & ME+HE & R \\
\hline
P0111605034 & 2017-12-29T02:10:24 & 29.94,\,-149.90 & (18.82,\,37.01) & ME & R \\
 & 2017-12-29T03:45:51 & 30.15,\,-173.94 & (19.00,\,37.23) & ME+HE & R \\
\hline
P0111605035 & 2017-12-31T05:04:28 & 36.29,\,160.69 & (24.22,\,43.57) & ME+HE & R \\
\hline
P0111605036 & 2018-01-02T01:36:37 & 41.21,\,-152.64 & (28.26,\,48.80) & ME+HE & R \\
 & 2018-01-02T03:12:03 & 41.37,\,-176.69 & (28.40,\,48.97) & ME+HE & R \\
\hline
P0111605037 & 2018-01-04T02:54:54 & 46.00,\,-178.26 & (32.07,\,54.05) & ME+HE & R \\
 & 2018-01-04T04:30:19 & 46.14,\,157.69 & (32.18,\,54.21) & ME+HE & R \\
\hline
P0111605040 & 2018-01-16T02:20:44 & 2.81,\,24.40 & (-7.35,\,12.30) & ME+HE & S \\
\hline
P0111605041 & 2018-01-19T06:50:09 & 35.81,\,-30.26 & (21.10,\,46.08) & LE+ME+HE & S \\
 & 2018-01-19T08:25:47 & 36.54,\,-53.76 & (21.69,\,46.83) & LE+ME+HE & S \\
\hline
P0111605042 & 2018-01-22T00:26:56 & 39.56,\,-153.64 & (24.19,\,50.11) & ME+HE & R \\
 & 2018-01-22T02:02:33 & 38.88,\,-177.12 & (23.62,\,49.36) & ME+HE & R \\
 & 2018-01-22T03:38:11 & 38.16,\,159.40 & (23.05,\,48.61) & ME+HE & R \\
\hline
P0111605043 & 2018-01-25T22:27:49 & -0.31,\,-108.09 & (-10.16,\,9.23) & ME+HE & R \\
\hline
P0111605046 & 2018-02-08T15:13:41 & 41.08,\,-172.55 & (28.15,\,48.66) & LE+ME+HE & S \\
 & 2018-02-08T16:49:07 & 40.91,\,163.41 & (28.03,\,48.47) & LE+ME+HE & S \\
 & 2018-02-08T18:24:33 & 40.74,\,139.36 & (27.88,\,48.30) & LE+ME+HE & S \\
\hline
P0111605047 & 2018-03-03T20:37:43 & 26.95,\,-128.96 & (16.23,\,33.98) & ME+HE & R \\
 & 2018-03-03T22:13:10 & 27.17,\,-153.00 & (16.42,\,34.20) & ME+HE & R \\
 & 2018-03-03T23:48:36 & 27.39,\,-177.04 & (16.61,\,34.42) & ME+HE & R \\
\hline
P0111605048 & 2018-02-13T03:23:20 & 28.06,\,-7.59 & (17.20,\,35.10) & LE+ME & S \\
 & 2018-02-13T04:58:47 & 27.85,\,-31.63 & (17.01,\,34.88) & LE+ME & S \\
\hline
P0111605050 & 2018-02-21T10:15:11 & -3.09,\,-132.42 & (-11.25,\,4.88) & ME+HE & S \\
\hline
P0111605051 & 2018-02-27T11:01:50 & -23.28,\,-158.80 & (-30.70,\,-13.35) & HE & S \\
\hline
P0111605053 & 2018-03-09T21:22:09 & 43.96,\,-157.11 & (30.46,\,51.80) & ME+HE & R \\
 & 2018-03-09T22:57:35 & 44.12,\,178.84 & (30.59,\,51.96) & ME+HE & R \\
 & 2018-03-10T00:33:01 & 44.27,\,154.79 & (30.70,\,52.13) & ME+HE & R \\
\hline
P0111605054 & 2018-03-13T03:17:49 & 50.69,\,104.14 & (35.62,\,59.43) & ME+HE & R \\
 & 2018-03-13T23:58:15 & 52.21,\,151.34 & (36.74,\,61.25) & ME+HE & R \\
 & 2018-03-14T01:33:40 & 52.32,\,127.27 & (36.82,\,61.39) & ME+HE & R \\
\hline
P0111605055 & 2018-03-20T16:05:51 & -12.37,\,109.20 & (-21.18,\,-2.29) & ME+HE & S \\
 & 2018-03-20T17:41:22 & -11.98,\,85.43 & (-20.82,\,-1.95) & ME+HE & S \\
\hline
P0111605058 & 2018-03-26T13:53:10 & 41.85,\,162.92 & (26.08,\,52.64) & ME+HE & S \\
 & 2018-03-26T15:28:47 & 42.47,\,139.45 & (26.60,\,53.34) & ME+HE & S \\
\hline
P0111605059 & 2018-03-28T18:44:51 & 38.78,\,-132.38 & (23.59,\,49.25) & ME+HE & R \\
\hline
P0111605060 & 2018-03-30T07:24:14 & 21.95,\,45.84 & (9.46,\,31.51) & ME+HE & R \\
 & 2018-03-30T08:59:51 & 21.24,\,22.21 & (8.84,\,30.74) & ME+HE & R \\
 & 2018-03-30T10:35:28 & 20.52,\,-1.42 & (8.23,\,30.03) & ME+HE & R \\
\hline
P0111605061 & 2018-04-01T02:25:13 & 4.15,\,126.50 & (-6.15,\,13.58) & ME & R \\
 & 2018-04-01T04:00:47 & 3.57,\,102.77 & (-6.67,\,13.02) & ME & R \\
 & 2018-04-01T05:36:22 & 2.98,\,79.05 & (-7.19,\,12.43) & ME & R \\
\hline
P0111605063 & 2018-08-30T08:19:01 & 24.10,\,81.11 & (13.72,\,31.11) & ME & S \\
 & 2018-08-30T09:54:27 & 23.87,\,57.07 & (13.52,\,30.88) & ME & S \\
 & 2018-08-30T13:05:17 & 23.40,\,9.00 & (13.12,\,30.41) & ME & S \\
 & 2018-08-30T16:16:09 & 22.94,\,-39.08 & (12.69,\,29.95) & ME+HE & S \\
 & 2018-08-31T01:48:48 & 21.53,\,176.69 & (11.44,\,28.55) & ME+HE & S \\
\end{longtable}
\end{center}
\end{document}